\documentclass[twocolumn]{aastex62}

\usepackage{longtable}
\usepackage{amssymb,amsmath}
\usepackage{array,booktabs}
\usepackage{graphicx}
\usepackage{multirow}
\usepackage{hyperref}
\usepackage{color}
\usepackage{soul}
\usepackage{rotating}

\definecolor{darkgreen}{rgb}{0.0,,0.0}

\newcommand{\HL}[1]{\textcolor{black}{#1}}

\shorttitle{}
\shortauthors{}

\begin{document}

\title{OGLE-2014-BLG-0962 and the First Statistical \HL{Validation of Bayesian Priors} for Galactic Microlensing}

\correspondingauthor{Yutong Shan}\email{yshan@cfa.harvard.edu}

\author{Yutong Shan}\affiliation{Harvard-Smithsonian Center for Astrophysics, 60 Garden Street, Cambridge, MA 02138, USA }

\author{Jennifer C. Yee}\altaffiliation{The {\emph{Spitzer}} Team}\affiliation{Harvard-Smithsonian Center for Astrophysics, 60 Garden Street, Cambridge, MA 02138, USA }

\author{Andrzej Udalski}\altaffiliation{The OGLE Collaboration}\affiliation{Warsaw University Observatory, Al. Ujazdowskie 4, 00-478 Warszawa, Poland}

\author{Ian A. Bond}\altaffiliation{The MOA Collaboration}\affiliation{Institute of Natural and Mathematical Sciences, Massey University, Auckland 0745, New Zealand}

\author{Yossi Shvartzvald}\altaffiliation{The Wise Group}\affiliation{Jet Propulsion Laboratory, California Institute of Technology, 4800 Oak Grove Drive, Pasadena, CA
91109, USA}

\author{In-Gu Shin}\affiliation{Harvard-Smithsonian Center for Astrophysics, 60 Garden Street, Cambridge, MA 02138, USA }

\author{Youn-Kil Jung}\affiliation{Harvard-Smithsonian Center for Astrophysics, 60 Garden Street, Cambridge, MA 02138, USA }

\nocollaboration


\author{Sebastiano Calchi Novati}\affiliation{
IPAC, Mail Code 100-22, Caltech, 1200 E. California Blvd., Pasadena, CA 91125, USA}

\author{Charles A. Beichman}\affiliation{NASA Exoplanet Science Institute, MS 100-22, California Institute of Technology, Pasadena, CA 91125, USA}

\author{Sean Carey}\affiliation{Spitzer Science Center, MS 220-6, California Institute of Technology, Pasadena, CA, USA}

\author{B. Scott Gaudi}\affiliation{Department of Astronomy, Ohio State University, 140 W. 18th Ave., Columbus, OH 43210, USA}

\author{Andrew Gould}\affiliation{Department of Astronomy, Ohio State University, 140 W. 18th Ave., Columbus, OH 43210, USA}

\author{Richard W. Pogge}\affiliation{Department of Astronomy, Ohio State University, 140 W. 18th Ave., Columbus, OH 43210, USA}

\collaboration{(The {\emph{Spitzer}} Team)}


\author{Rados{\l}aw Poleski}\affiliation{Warsaw University Observatory, Al.~Ujazdowskie~4, 00-478~Warszawa,
Poland}\affiliation{Department of Astronomy, Ohio State University, 140 W. 18th Ave., Columbus, OH  43210, USA}

\author{Jan Skowron}\affiliation{Warsaw University Observatory, Al.~Ujazdowskie~4, 00-478~Warszawa,
Poland}

\author{Szymon Koz{\l}owski}\affiliation{Warsaw University Observatory, Al.~Ujazdowskie~4, 00-478~Warszawa,
Poland}

\author{Przemys{\l}aw Mr{\'o}z}\affiliation{Warsaw University Observatory, Al. Ujazdowskie 4, 00-478 Warszawa, Poland}

\author{Pawe{\l} Pietrukowicz}\affiliation{Warsaw University Observatory, Al. Ujazdowskie 4, 00-478 Warszawa, Poland}


\author{Micha{\l} K. Szyma{\'n}ski}\affiliation{Warsaw University Observatory, Al. Ujazdowskie 4, 00-478 Warszawa, Poland}

\author{Igor Soszy{\'n}ski}\affiliation{Warsaw University Observatory, Al. Ujazdowskie 4, 00-478 Warszawa, Poland}

\author{Krzysztof Ulaczyk}\affiliation{Warsaw University Observatory, Al. Ujazdowskie 4, 00-478 Warszawa, Poland}

\author{{\L}ukasz Wyrzykowski}\affiliation{Warsaw University Observatory, Al. Ujazdowskie 4, 00-478 Warszawa, Poland}

\collaboration{(The OGLE Collaboration)}


\author{Fumio Abe}\affiliation{Institute for Space-Earth Environmental Research, Nagoya University, Nagoya 464-8601, Japan}

\author{Richard K. Barry}\affiliation{Code 667, NASA Goddard Space Flight Center, Greenbelt, MD 20771, USA}

\author{David P. Bennett}\affiliation{Code 667, NASA Goddard Space Flight Center, Greenbelt, MD 20771, USA}\affiliation{Department of Astronomy, University of Maryland, College Park, MD 20742, USA}

\author{Aparna Bhattacharya}\affiliation{Code 667, NASA Goddard Space Flight Center, Greenbelt, MD 20771, USA}\affiliation{Department of Astronomy, University of Maryland, College Park, MD 20742, USA}


\author{Martin Donachie}\affiliation{Department of Physics, University of Auckland, Private Bag 92019, Auckland, New Zealand}

\author{Akihiko Fukui}\affiliation{Okayama Astrophysical Observatory, National Astronomical Observatory of Japan, 3037-5 Honjo, Kamogata, Asakuchi, Okayama 719-0232, Japan}

\author{Yuki Hirao}\affiliation{Department of Earth and Space Science, Graduate School of Science, Osaka University, Toyonaka, Osaka 560-0043, Japan}

\author{Yoshitaka Itow}\affiliation{Institute for Space-Earth Environmental Research, Nagoya University, Nagoya 464-8601, Japan}

\author{Kohei Kawasaki}\affiliation{Department of Earth and Space Science, Graduate School of Science, Osaka University, Toyonaka, Osaka 560-0043, Japan}

\author{Iona Kondo}\affiliation{Department of Earth and Space Science, Graduate School of Science, Osaka University, Toyonaka, Osaka 560-0043, Japan}



\author{Naoki Koshimoto}\affiliation{Department of Astronomy, Graduate School of Science, The University of Tokyo, 7-3-1 Hongo, Bunkyo-ku, Tokyo 113-0033, Japan
National Astronomical Observatory of Japan, 2-21-1 Osawa, Mitaka, Tokyo 181-8588, Japan}

\author{Man Cheung Alex Li}\affiliation{Department of Physics, University of Auckland, Private Bag 92019, Auckland, New Zealand}

\author{Yutaka Matsubara}\affiliation{Institute for Space-Earth Environmental Research, Nagoya University, Nagoya 464-8601, Japan}

\author{Yasushi Muraki}\affiliation{Institute for Space-Earth Environmental Research, Nagoya University, Nagoya 464-8601, Japan}

\author{Shota Miyazaki}\affiliation{Department of Earth and Space Science, Graduate School of Science, Osaka University, Toyonaka, Osaka 560-0043, Japan}

\author{Masayuki Nagakane}\affiliation{Department of Earth and Space Science, Graduate School of Science, Osaka University, Toyonaka, Osaka 560-0043, Japan}

\author{Cl{\'e}ment Ranc}\affiliation{Code 667, NASA Goddard Space Flight Center, Greenbelt, MD 20771, USA}

\author{Nicholas J. Rattenbury}\affiliation{Department of Physics, University of Auckland, Private Bag 92019, Auckland, New Zealand}

\author{Haruno Suematsu}\affiliation{Department of Earth and Space Science, Graduate School of Science, Osaka University, Toyonaka, Osaka 560-0043, Japan}

\author{Denis J. Sullivan}\affiliation{School of Chemical and Physical Sciences, Victoria University, Wellington, New Zealand}


\author{Takahiro Sumi}\affiliation{Department of Earth and Space Science, Graduate School of Science, Osaka University, Toyonaka, Osaka 560-0043, Japan}

\author{Daisuke Suzuki}\affiliation{Institute of Space and Astronautical Science, Japan Aerospace Exploration Agency, 3-1-1 Yoshinodai, Chuo, Sagamihara, Kanagawa, 252-5210, Japan}

\author{Paul J. Tristram}\affiliation{ University of Canterbury Mt.\ John Observatory, P.O. Box 56, Lake Tekapo 8770, New Zealand}


\author{Atsunori Yonehara}\affiliation{Department of Physics, Faculty of Science, Kyoto Sangyo University, 603-8555 Kyoto, Japan}

\collaboration{(The MOA Collaboration)}


\author{Dan Maoz}\affiliation{School of Physics and Astronomy and Wise Observatory, Tel-Aviv University, Tel-Aviv 6997801, Israel}

\author{Shai Kaspi}\affiliation{School of Physics and Astronomy and Wise Observatory, Tel-Aviv University, Tel-Aviv 6997801, Israel}

\author{Matan Friedmann}\affiliation{School of Physics and Astronomy and Wise Observatory, Tel-Aviv University, Tel-Aviv 6997801, Israel}

\collaboration{(The Wise Group)} 


\begin{abstract}

OGLE-2014-BLG-0962 (OB140962) is a stellar binary microlensing event that was well-covered by observations from the {\emph{Spitzer}} satellite as well as ground-based surveys. Modelling yields a unique physical solution: a mid-M+M-dwarf binary with $M_{\rm prim} = 0.20 \pm 0.01 M_\sun$ and \HL{$M_{\rm sec} = 0.16 \pm 0.01 M_\sun$}, with projected separation of $2.0 \pm 0.3$ AU. The lens is only $D_{\rm LS} = 0.41 \pm 0.06$ kpc in front of the source, making OB140962 a bulge lens and the most distant {\emph{Spitzer}} binary lens to date. \HL{In contrast,} because the Einstein radius ($\theta_{\rm E} = 0.143 \pm 0.007$ mas) is unusually small, a standard Bayesian analysis, conducted in the absence of parallax information, would predict a brown dwarf binary. We test the accuracy of Bayesian analysis over a set of {\emph{Spitzer}} lenses, \HL{finding overall good agreement throughout the sample.} We also illustrate the methodology for probing the Galactic distribution of planets by comparing the cumulative distance distribution of the {\emph{Spitzer}} 2-body lenses to that of the {\emph{Spitzer}} single lenses. 

\bigskip

\keywords{gravitational lensing: micro -- binaries: general -- stars: low-mass -- Galaxy: bulge -- methods: statistical}  
\end{abstract}

\section{Introduction}\label{s:intro}

Gravitational microlensing is a type of transient phenomenon in which a temporary alignment of a foreground lens and a background source causes the source to be magnified. Because lensing is sensitive to the presence of mass independent of any associated flux, it represents a unique means to probe distant and faint populations of many compact astrophysical objects of interest (e.g. low-mass main-sequence stars, brown dwarfs, planets, and stellar remnants). Indeed, microlensing has discovered stellar remnants \citep[e.g.][]{Shvartzvald15,Wyrzykowski16} and free floating planets \citep[e.g.][]{Sumi11, Mroz17, Mroz18}, as well as characterized planets and low-mass objects anywhere between the Sun and the Galactic center. 

One drawback of the microlensing technique is that, while relative parameters such as mass ratios (for binaries) are routinely measured, it is often difficult to infer the absolute physical properties of the lens from a ground-based light curve alone. For many applications, the absolute physical properties of the lens -- its mass ($M_L$), distance ($D_L$), and lens-source relative kinematics ($\vec{v}_{\rm rel}$) -- are paramount to interpretation. For example, in characterizing individual lensing systems, incertitude in the physical parameters can make the difference between a star, a brown dwarf, a planet, or a moon \citep[e.g.][]{Bennett14,Albrow18}. To study the distribution of planet masses from an ensemble of binary lens mass-ratios \citep{Shvartzvald16b,Suzuki16}, the host masses are needed.  Measurement of planetary occurrence rate as a function of distance and Galactic environment \citep[e.g.][]{Penny16} also depends on whether the system distances and kinematic memberships are reliably assigned on average.       

The challenge of determining physical quantities for lenses from ground-based data alone arises because 4 parameters are needed to constrain the 4 physical properties (that is, 2 scalar quantities: $M_L$ and $D_L$, and 1 vector quantity with 2 components: $\vec{v}_{\rm rel} = [v_{{\rm rel},N}, v_{{\rm rel},E}$]). The microlensing parameters $t_{\rm E}$ (Einstein timescale), $\theta_{\rm E}$ (Einstein radius), and $\vec{\pi}_{\rm E}$ (microlensing parallax) form a complete set that can be solved for the physical parameters (see Section \ref{ss:ulensing-pars}). However, only $t_{\rm E}$ is readily measured for microlensing events. One can measure $\theta_{\rm E}$ using the finite-source effect, which is not usually present for single-lens events but often feasible for binary lensing. This leaves $\vec{\pi}_{\rm E}$ which, prior to {\emph{Spitzer}}, was not accessible for most events because their $t_{\rm E}$ is considerably less than 1 year.   

In most cases for which $\vec{\pi}_{\rm E}$ is not available, a Bayesian analysis can be used to obtain a posterior on the physical properties of a particular event. This is done by forward modelling individual source and lens stars in the Galaxy to match with the measured $t_{\rm E}$ and $\theta_{\rm E}$ of the event. The priors for such a model integrate kinematics, stellar density profiles, and mass functions for the Galactic disk and bulge. Since most microlensing events will have ground-based survey data only, Bayesian analysis will continue to be the leading avenue used to estimate the physical parameters of microlensing systems, until the {\emph{WFIRST}} \citep{Spergel13} era.  

Given the importance of physical parameters for the correct interpretation of lensing systems, it is critical to examine the accuracy of Bayesian analysis -- that is, to compare its predictions to the `true' values determined from other means. One way to arrive at the `true' answers is to perform followup adaptive optics (AO) imaging, specifically to resolve the source and lens separately. Notably, AO solutions were obtained in a handful of cases, finding generally good agreement with the original Bayesian predictions. For example, \citet{Batista14} found the host mass and distance of MOA-2011-BLG-293 to be $M_L = 0.86 \pm 0.06 M_\sun$ and $D_L = 7.72 \pm 0.44$ kpc, fully consistent with the Bayesian predictions of $M_L=0.59^{+0.35}_{-0.29} M_\sun$ and $D_L=7.15\pm0.75$ kpc \HL{\citep{Yee12} }. For 
OGLE-2005-BLG-169, \citet{Bennett15} and \citet{Batista15} retrieve physical properties ($M_L = 0.69 \pm 0.02 M_\sun$ and $D_L = 4.1 \pm 0.4$ kpc) consistent with the original Bayesian result from \citet{Gould06} ($M_L = 0.49^{+0.23}_{-0.29} M_\sun$ and $D_L = 2.7^{+1.6}_{-1.3}$kpc).    

Simultaneous observation by a distant satellite provides another way to obtain the `true' parameters of a microlensing event. Since microlensing events involve very precise alignment between the lens system and the source star trajectory, the alignment angle is different for two widely separated observers, leading to inter-light curve discrepancies that can be modelled to yield $\vec{\pi}_{\rm E}$. This principle motivated the {\emph{Spitzer}} microlensing campaign which, since its inaugural year in 2014, has yielded numerous parallax measurements to single lenses \citep[e.g.][]{CalchiNovati15a,Yee15b,Zhu17a}. Moreover, {\emph{Spitzer}} has helped to measure $\theta_{\rm E}$ and $\pi_{\rm E}$ for a dozen stellar binaries \citep[e.g.][]{Bozza16,Han17,Wang17} and planets \citep[e.g.][]{Udalski15b,Street16,Shvartzvald17,Ryu18}, disentangling the absolute physical properties of the lens systems. 

The {\emph{Spitzer}} sample also forms an ideal test bed for Bayesian analysis. At least two {\emph{Spitzer}} objects have published Bayesian analysis results. The physical properties of the single-lens OB151482 found by detailed modelling of the finite source effect \citep{Chung17} yielded a mass and distance consistent with a Bayesian analysis presented by \citet{Zhu17a} for the same object. However, for the low mass-ratio planet OB161195, \citet{Bond17}'s Bayesian predictions are \HL{in tension} with the parameters found by a full modelling including the {\emph{Spitzer}} light curve presented in \citet{Shvartzvald17}. 

The variation in the results for AO and parallax tests indicate the need for a systematic test of the Bayesian method as compared to true measurements. This work presents the discovery of OGLE-2014-BLG-0962 (OB140962), a textbook example of a stellar binary lens with excellent data coverage from both the ground and {\emph{Spitzer}}, leading to superbly constrained physical parameters. We describe the data in Section \ref{s:obs} and the modelling process in Section \ref{s:ana}. In particular, we give the mathematical relations between the microlensing and physical parameters of interest in Section \ref{ss:ulensing-pars}. The unique physical properties for this lens are given in Section \ref{s:res}, where we find that this binary is likely the most distant stellar binary detected by {\emph{Spitzer}} to date, almost certainly a member of the Galactic bulge. 

In Section \ref{s:bayes} we start by performing Bayesian analysis on OB140962 while withholding the parallax information (Section \ref{ss:bayes-0962}). We subsequently repeat this analysis for {\emph{Spitzer}} events with secure parallax-derived physical parameters to investigate the overall reliability of Bayesian analysis. 

Finally, in Section \ref{s:binarity}, we compare the distance distribution of well-characterized {\emph{Spitzer}} binaries (including planets) to that of the {\emph{Spitzer}} single lenses. This serves to illustrate how one might quantify the relative occurrence rate of planets and binaries throughout the Galaxy when a larger sample becomes available and selection effects are systematically quantified. Section \ref{s:sum} provides a summary.

\bigskip
\section{Observations}\label{s:obs}

\begin{figure*}[htb]
\centering
\includegraphics[scale=1]{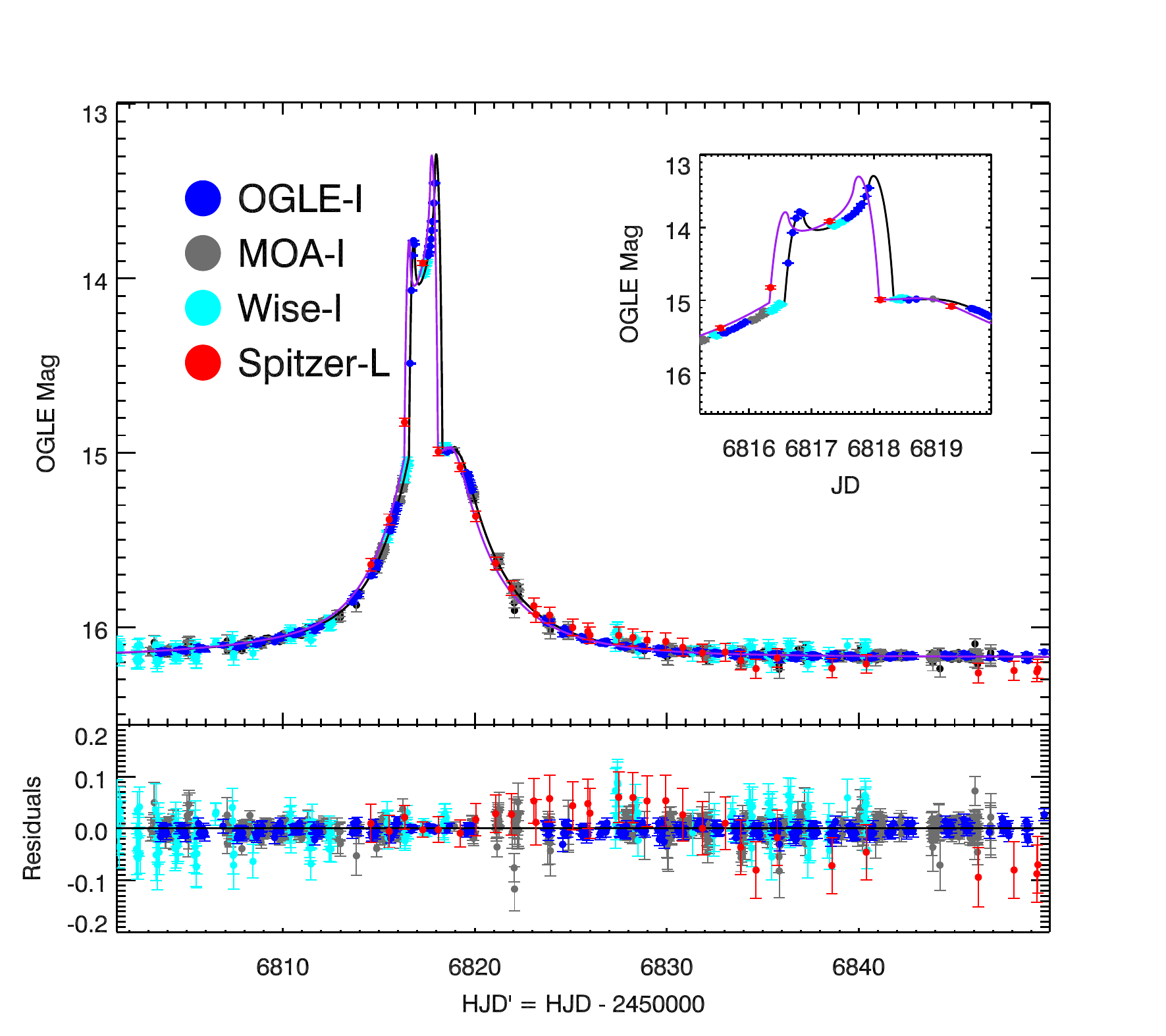}
\caption{The ground- and space-based light curves of binary microlensing event OGLE-2014-BLG-0962, including data from OGLE, MOA, Wise, and {\emph{Spitzer}}. The best-fit $u_0 > 0$ model is shown in  black, while the purple curve outlines the corresponding space-based solution. The inset shows a zoom-in of the highly magnified portion.}
\label{f:lc}
\end{figure*}

OGLE-2014-BLG-0962 (hereafter OB140962) was located at equatorial coordinates $(\alpha, \delta)_{\rm J2000}=$ (18:01:42.98,-27:55:56.2). These translate into Galactic coordinates $(l,b)=(2.7^\circ,-2.5^\circ)$. It was alerted by the Optical Gravitational Lensing Experiment Early Warning System (OGLE: \citealt{Udalski15a}; EWS: \citep{Udalski03}) at UT 18:53, 30 May 2014, in time to mobilize immediate followup observations by the first {\emph{Spitzer}} microlensing campaign \citep[e.g.][]{Udalski15b,Yee15a}. The light curve, as shown in Figure \ref{f:lc}, is a caustic crossing event that reveals a clear signature of a high mass-ratio binary lens. The caustic entrance and exit are well covered by both ground- (\textsection \ref{ss:ground-obs}) and space-based (\textsection \ref{ss:space-obs}) observations, leading to secure determination of microlensing and physical lens parameters. Prior to modelling, the errors on the photometric reduction from each observatory were rescaled according to standard procedures and clipped for outliers (\textsection \ref{ss:errors}).      

\bigskip
\subsection{Ground-based Observations}\label{ss:ground-obs}

The OGLE observations are conducted with the 1.3m Warsaw telescope at the Las Campanas Observatory in Chile, with a $1.4 ~{\rm deg}^2$ field of view (FoV) camera. Data in the $I$-band were taken at a nominal cadence of $\Gamma=1\,{\rm hr}^{-1}$. The regions around and between the caustic crossings are well-sampled. $V$-band data were also acquired at a lower cadence. Four points, including one near peak, were captured in OGLE $V$-band between HJD' = 6810 and 6830. OGLE photometry was reduced with the Difference-Imaging Analysis (DIA) method \citep{Alard98,Wozniak00}. 

The Microlensing Observations in Astrophysics (MOA) collaboration also observed this event (MOA-2014-BLG-285). MOA provided an alert on 31 May 2014. The MOA survey is conducted from the University of Canterbury Mt John Observatory in New Zealand, which features a 1.8 m telescope with a camera whose FoV is 2.2 ${\rm deg}^2$. MOA observes in the custom $R_{\rm MOA}$-band, which is approximately the superposition of the standard $I$ and $R$-bands. Normally MOA surveys at high cadence, though for this particular event it missed the portion of the light curve between the caustic entrance and exit due to weather.  MOA photometry is reduced by the DIA pipeline summarized in \citet{Bond01}.     

\HL{The Wise microlensing survey \citep{Shvartzvald16b} used the 1 m telescope at Tel Aviv University's Wise Observatory in Israel. The Large Area Imager for the Wise Observatory camera with (FoV = 1 ${\rm deg}^2$) was used to collect data on this event in survey mode.} Wise observations are conducted in the $I$-filter. The nominal cadence is \HL{30 min, averaging $\sim 5$ to $6$ observations per night due to target visibility.} Photometry for Wise is performed with the DIA software described in \citet{Albrow09}.

\bigskip
\subsection{Space-based Observations}\label{ss:space-obs}

OB140962 was observed in the first season of the {\emph{Spitzer}} microlensing campaign, before the development of the objective target selection procedure of \citet{Yee15a}, which was used in subsequent seasons. {\emph{Spitzer}} observations were taken in the $L$-band ($3.6 \mu m$). The target was selected for Spitzer observations before it showed any features attributable to a binary. Observations began on 6 June 2014 (HJD' = 6814.59). The last data point was taken on 10 July 2014 (HJD' = 6849.34). A total of 31 observations were taken at a cadence of $\Gamma \sim 1 ~{\rm day}^{-1}$. It captured several points during the anomaly. {\emph{Spitzer}} photometry was reduced with the pipeline presented in \citet{CalchiNovati15b}.

\bigskip
\subsection{Error Rescaling}\label{ss:errors}

Purely photometric (i.e., Poisson) errors are often underestimated, which prompts a rescaling of the data points and renormalization of the errors. Except for the OGLE errors, which are rescaled based on the recommended procedure of \citet{Skowron16}, errors from the other observatories (in magnitudes) are modified using the scheme of \citet{Yee12}\footnote{The original formulation is  $\sigma_{\rm rescaled}=k\sqrt{\sigma_{\rm pipeline}^2+\sigma_{\rm min}^2}$, though in many cases including this one, $\sigma_{\rm min} \approx 0$}, where: 

\begin{equation}
\sigma_{\rm rescaled}=k\sigma_{\rm pipeline}.      
\label{e:error-scale}
\end{equation}

With $\sigma_{\rm rescaled}$, every data point should contribute $\sim1$ in $\chi^2$ on average. The $k$-factor is adjusted manually until this is the case. The newly scaled errors are used to reject outliers and updated iteratively. As a result, we reject four OGLE, two MOA, and two Wise measurements. The final adopted $k$ parameters are listed in Table \ref{t:error-scale}.  

\setlength{\tabcolsep}{10pt}
\begin{table}[tbh]
\begin{center} 
\caption{Error Rescaling Factors for Each Observatory}
\label{t:error-scale}
\begin{tabular}{c|c}
\hline\hline
Observatory  &  $k$ \\
\hline
OGLE & See \citet{Skowron16} \\
MOA &  1.25 \\
Wise & 1.46 \\
{\emph{Spitzer}} & 7.9 \\
\hline\hline
\end{tabular}
\end{center}
\end{table}

\bigskip
\section{Analysis}\label{s:ana}

In this section we deduce the microlensing parameters and physical lens properties via joint modelling of the ground- and space-based light curves. In \textsection \ref{ss:ulensing-pars} we introduce the relevant parameters that can be measured from our data and describe how to use them to obtain the absolute physical properties. But even prior to rigorous modelling, many conclusions can be drawn from the data points alone thanks to the comprehensive coverage of the event from both the ground and space. Therefore, in \textsection \ref{ss:heuristic} we provide a heuristic description of the light curve, which yields basic insight into the nature of the microlensing event. \textsection \ref{ss:modelling} summarizes the multi-staged modelling process to eventually arrive at the microlensing parameters. 

\bigskip
\subsection{Microlensing Parameters \& Relations to Physical Properties}\label{ss:ulensing-pars}

There are six fundamental parameters associated with and routinely measured for binary lensing light curves: ($t_0$, $u_0$, $t_{\rm E}$, $s$, $q$, $\alpha$). The first three quantities stand for the peak time, impact parameter of the source trajectory to the lens (scaled to the lens Einstein radius, $\theta_{\rm E}$, see below), \HL{and} the Einstein time scale, i.e., the characteristic width of the portion of the light curve undergoing magnification. The second set of 3 parameters pertain to the binary lens. They are: the instantaneous projected separation between the components (normalized to $\theta_{\rm E}$), their mass ratio, and the projected angle of the source trajectory to the binary axis, respectively. 

All the parameters presented so far are either geometric or relative. Of course, it is the absolute physical properties of the lens that are of greatest interest. The lens mass ($M_L$), distance ($D_L$), and relative proper motion between the lens and source ($\mu_{\rm rel}$) can be determined provided additional effects are measured. They are linked to the direct observables via the Einstein radius ($\theta_{\rm E}$) and the dimensionless vector microlensing parallax ($\vec{\pi}_{\rm E}$). For $\pi_{\rm rel} \equiv \pi_L - \pi_S = {\rm AU}(D_L^{-1} - D_S^{-1})$, $D_S$ being the source distance (usually close to the Galactic center at $\sim 8.3$ kpc), $\theta_{\rm E}$ and $\vec{\pi}_{\rm E}$ are defined as follows \citep[e.g.][]{Gould00}:
\begin{equation}
\theta_{\rm E} \equiv \sqrt{\kappa M_L \pi_{\rm rel}}; ~~~~~ \kappa \equiv \frac{4G}{c^2 {\rm AU}} \approx 8.14 \frac{\rm mas}{M_\sun};
\label{e:thetae}
\end{equation}

\noindent and 
\begin{equation}
\vec{\pi}_{\rm E} \equiv \frac{\pi_{\rm rel}}{\theta_{\rm E}} \frac{\vec{\mu}_{\rm rel}}{|\mu_{\rm rel}|}; ~~~~~ \mu_{\rm rel} = \frac{\theta_{\rm E}}{t_{\rm E}}.
\label{e:pie}
\end{equation}

\noindent Manipulating Equations (\ref{e:thetae}) and (\ref{e:pie}), both the lens mass and distance can be expressed as a function of $\theta_{\rm E}$ and $\pi_{\rm E}$: 

\begin{equation}
M_L = \frac{\theta_{\rm E}}{\kappa \pi_{\rm E}}; ~~~~~ D_L = \frac{\rm AU}{\pi_{\rm E}\theta_{\rm E}+{\rm AU}/D_S}.
\label{e:ml}
\end{equation}

For caustic-crossing events, the finite source effect constrains a 7th parameter $\rho$, where

\begin{equation}
\rho \equiv \frac{\theta_{\rm \star}}{\theta_{\rm E}} = \frac{t_\star}{t_{\rm E}},
\label{e:rho}
\end{equation}

\noindent i.e. the size of the source $\theta_\star$ measured in units of $\theta_{\rm E}$. If the source radius can be deduced independently, for instance from the event's position in the local color-magnitude diagram (CMD), then $\theta_{\rm E}$ can be calculated. Alternatively, $\rho$ can be defined as the source self-crossing time, $t_*$, relative to $t_E$.


The microlensing parallax, $\vec{\pi}_{\rm E}$, can be measured with a second line of sight to the event, which generally will result in a light curve with timing and morphology that are distinct from the first because of the apparent difference in trajectory. A useful qualitative approximation for the components of $\vec{\pi}_{\rm E}$ is given by the scaled difference in the $t_0$ and $u_0$ between the two sight lines \citep[e.g.][]{Refsdal66}: 

\begin{equation}
\vec{\pi}_{\rm E} = \frac{\rm AU}{D_\bot}\left(\frac{\Delta t_0}{t_{\rm E}}; \Delta u_0 \right)
\label{e:pie-space}
\end{equation}

\noindent \HL{where $\vec{D}_\bot$ is the projected separation vector between the two observing locations in the plane of the sky. Equation (\ref{e:pie-space}) applies for the coordinate system in which the x-axis is aligned with $\vec{D}_\bot$.} 
For Earth and the {\emph{Spitzer}} satellite, the magnitude of this vector is approximately $1$ to $1.5$ AU. Refer to, e.g., Equations (8) to (10) in \citet{CalchiNovati15a} for the exact relation between $\vec{\pi}_{\rm E}$, $\Delta t_0$, $\Delta u_0$, and instantaneous $\vec{D}_\bot$, which is used in the actual modelling for $\vec{\pi}_{\rm E}$. 
    
\bigskip

\subsection{Heuristic Description of the Light Curve}\label{ss:heuristic}

The ground-based light curve has a broad double-horned structure characteristic of roughly equal mass-ratio binary lensing, exhibiting a clear caustic entrance (HJD' = HJD-2450000 $\sim 6816.6$) and exit (HJD' $\sim 6818.3$). The hump in the light curve immediately following the caustic exit (HJD' $\sim 6819.0$) implies a cusp approach. The shape and timing of these features place tight constraints on the event geometry. Below we illustrate the back-of-the-envelope process of converting the lightcurve components into microlensing parameters and interpret the inferred physical properties of the source-lens system.   

\begin{figure}[htb]
\hspace{-1.5cm}
\includegraphics[scale=0.6]{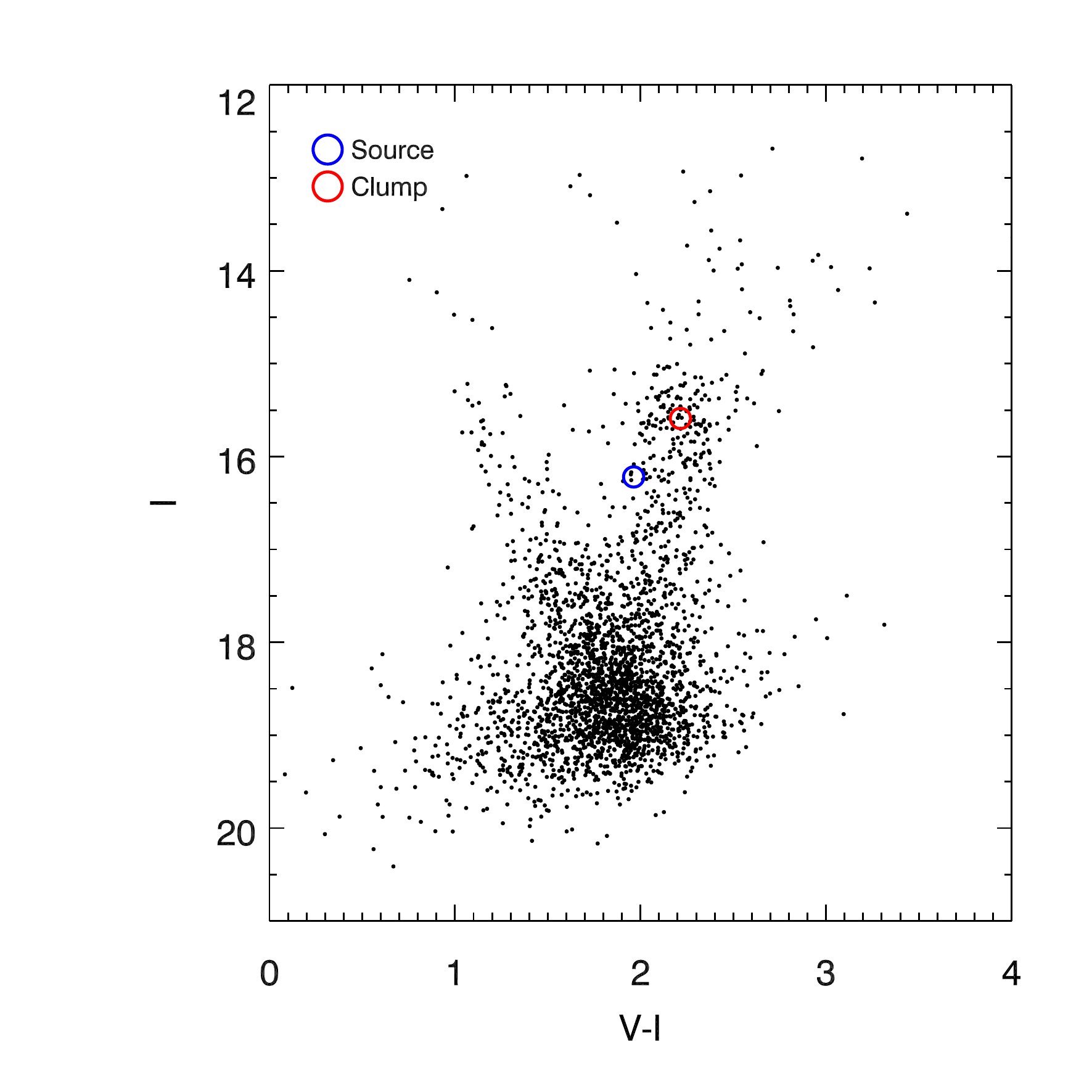}
\caption{The local color-magnitude diagram around OB140962. The red giant clump centroid is located at the center of the red circle. }
\vspace{+0.1cm}
\label{f:cmd}
\end{figure}

We approximate this event to have zero blending and estimate the Einstein timescale ($t_{\rm E}$) from the half-width of the magnified portion of the light curve at $1.3\times$ the baseline flux. For this event, $t_{\rm E} \sim 5$ days. The caustic crossings resolve the source size (i.e., finite source effect), allowing us to determine $\theta_{\rm E}$ from Equation (\ref{e:rho}). Figure \ref{f:lc} shows the half-width of the caustic entrance is $t_\star \sim 0.15$ days, thus $\rho \approx 0.03$. To obtain $\theta_\star$, we note that the event placement in the local color-magnitude diagram is consistent with a bulge giant \HL{(Figure \ref{f:cmd} and Section \ref{ss:sourcestar})}. A typical clump giant might have radius $5$ to $10\times$ that of the Sun, say $\theta_\star \approx 4 ~\mu {\rm as}$. Then, $\theta_{\rm E} \approx 0.13 ~{\rm mas}$. Together with $t_{\rm E}$ and Equation (\ref{e:pie}), we find that the relative lens-source proper motion is $\mu_{\rm rel} \approx 9.5 ~{\rm mas}/{\rm yr}$. 

For many microlensing discoveries with only ground-based data, deducing $\theta_{\rm E}$ and $\mu_{\rm rel}$ is as far as we can go. To estimate the absolute physical properties of the lens, we might assume the lens has a typical distance of $D_L \sim 6$ kpc. Assuming a source distance of $D_S \sim 8.3$ kpc, $\pi_{\rm rel} \approx 0.046$. Then, substituting $\theta_E \approx 0.13$ mas into Equation (\ref{e:thetae}) gives $M_{\rm tot} \approx 0.045 M_\sun$. Therefore, the atypically small $\theta_E$ for this event (\HL{normally $\sim 0.5$ mas}) implies an exciting low-mass BD-BD binary.


For this event we have parallax information, which constrains the true lens distance and mass. To estimate $\pi_{\rm E}$, we see from Equation (\ref{e:pie-space}) that $|\pi_{\rm E}| \approx \sqrt{(\Delta t_0/t_{\rm E})^2+(\Delta u_0)^2}$, where we have used the fact that $D_\bot$ is of order 1 AU. For OB140962, the {\emph{Spitzer}} light curve actually closely mimics the ground-based one in shape as well as timing --- offset by $\Delta t_0 \sim 0.3$ days. The virtually indistinguishable light curve morphologies between the two sightlines strongly implies nearly identical impact parameters between the two events (i.e.,  $\Delta u_0$ negligible), so $\pi_{\rm E} \sim \Delta t_0/t_{\rm E} \approx 0.06$. The physical parameters $M_{\rm tot}$ and $D_L$ can both be computed from $\theta_{\rm E}$ and $\pi_{\rm E}$. According to Equation (\ref{e:ml}), $M_{\rm tot} \approx 0.27 M_\sun$ and $D_L \approx 7.8 ~{\rm kpc}$. Here we have assumed that the source is located at the Galactocentric distance $R_0 \sim 8.3$ kpc. Therefore, from this heuristic evaluation of the ground light curves in conjunction with the {\emph{Spitzer}} parallax, we reach the conclusion that the lens is a typical low-mass binary that must be very close to the source. This is at odds with the earlier expectation of a very low-mass lens from $\theta_{\rm E}$ and  $\mu_{\rm rel}$ alone.   

\bigskip
\subsection{Modelling the Light Curve}\label{ss:modelling}

To map the overall topology of the parameter space for the ground-based light curve, we perform a grid search in $\chi^2$-space over $\log s$, $\log q$, and $\alpha$, which are responsible for the magnification profile of the event \HL{\citep{Dong06}}. For each point $(\log s \in [-1,1], \log q \in [-5, 1], \alpha \in [0, 2\pi])$ on the $100 \times 100 \times 21$ grid, we allow the other light curve parameters ($t_0$, $u_0$, $t_{\rm E}$, $\rho$) to be explored by a Markov Chain Monte Carlo (MCMC) until it settles on the minimal $\chi^2$. We find the global minimum $\chi^2$ to be in the region around $\log s \sim 0.3$ and $\log q \sim 0.07$ (based on modelling the OGLE $I$-band data prior to error rescaling). 

From the best-fit grid point we launch our full MCMC for a joint fit for all four data sets (3 ground, 1 space). \HL{The finite source effect is modelled using the ray-shooting method \citep{Schneider86,Kayser86,Wambsganss97} and regions of the light curve immediately adjacent to caustic crossings are computed through the hexadecapole approximation \citep{Gould08,Pejcha09}.} With the inclusion of the space data, two additional parameters associated with the space parallax are fit: $\vec{\pi}_{\rm E} = (\pi_{\rm E,N}, \pi_{\rm E,E})$. We modelled the limb-darkening of the source star using the parameters derived in \textsection \ref{ss:sourcestar}. The final best-fit solutions are compiled in Table \ref{t:bestpars}. The parameter errors presented are 16\% and 84\% confidence intervals (CIs), evaluated from the MCMC posteriors. 

\begin{figure}[htb]
\hspace{-0.5cm}
\includegraphics[scale=0.52]{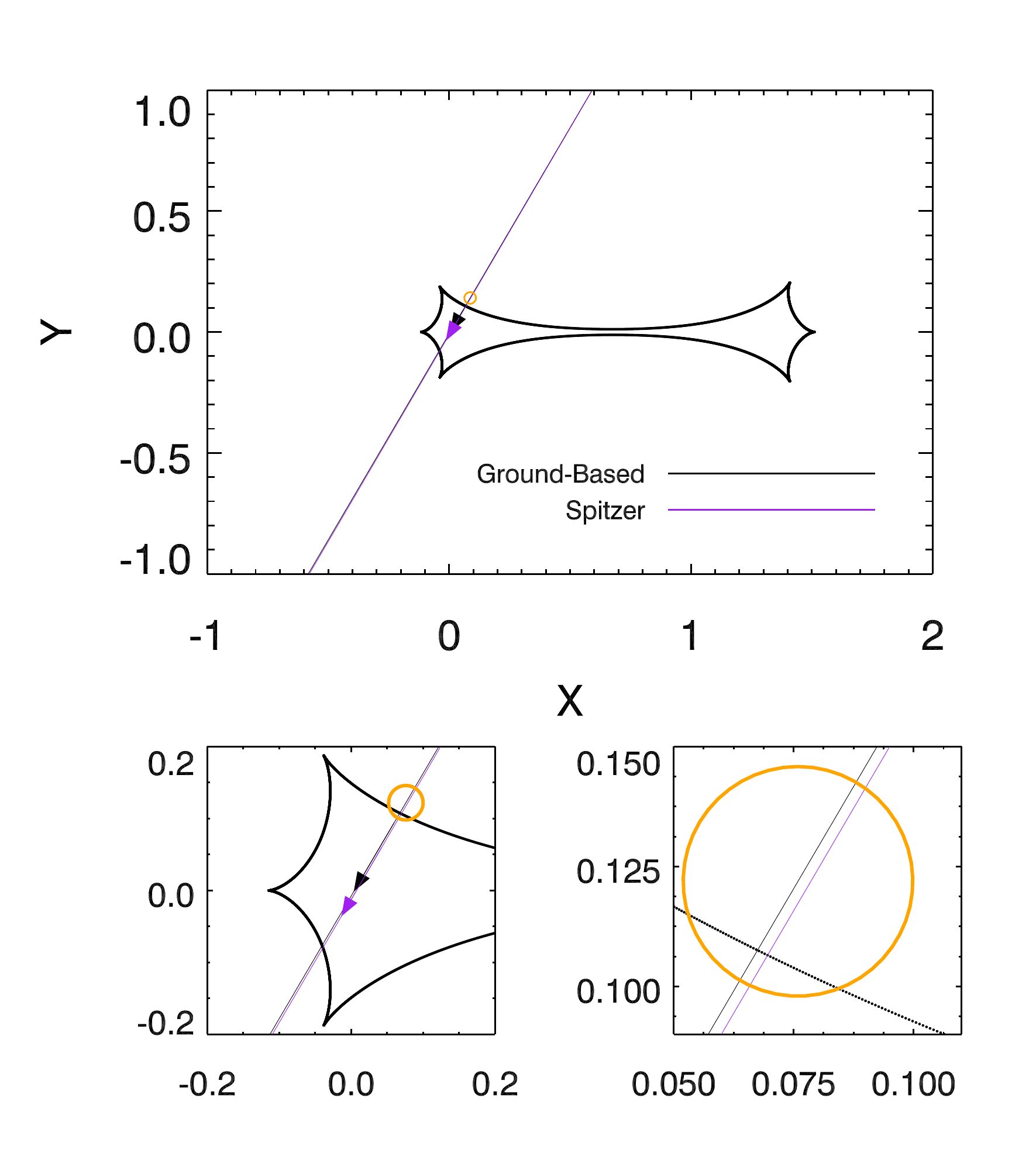}
\caption{The $u_0 > 0$ caustic structure of the binary microlensing event OB140962. This is a resonant caustic crossing event. The source size is indicated by the orange circle. Lower panels show zoom-ins of the ground (black)- and space (purple)-based source trajectories, with miniscule separation between them. Arrow tips coincide with $t_0$ for the ground-based light curve. }
\vspace{+0.1cm}
\label{f:caustics}
\end{figure}

Single-lens satellite parallax suffers from the well-known 4-fold degeneracy \citep[e.g.][]{Refsdal66,Gould94}. For binary lensing, depending on the data quality and coverage, some degeneracies can be resolved \citep[see discussion in, e.g.,][]{Zhu15}. OB140962 has a well-covered light curve from both the ground and space, leading to a straightforward interpretation. In this case, the $u_0 > 0$ and $u_0 < 0$ degeneracy persists, but maps into nearly identical physical properties. Therefore, the physical solution is unique. In Figure \ref{f:lc} we plot the $u_0 > 0$ solution based on the corresponding best-fitting parameters. Figure \ref{f:caustics} shows the associated caustic structure. 
\setlength{\tabcolsep}{4pt}
\begin{table*}[htb]
\begin{center}
\caption{Posterior and Best-Fit Microlensing Parameters Combining Ground-Space Observations}
\label{t:bestpars}
\begin{tabular}{lrrrrrrrrrrrr}
\hline\hline
Model & $\chi^2_{\rm{total}}$ & $t_0 - 6817 $& $u_0$& $t_E$& $\log (s)$& $\log (q)$& $\alpha$& $\rho$& $\pi_{\rm{E},N}$& $\pi_{\rm{E},E}$& $f_{s, OGLE}$& $f_{b, OGLE}$ \\
& & (HJD') & & (days) & & & (rad) & & & & & \\
\hline
              $u_0 > 0$ Median  &                           ...  &    0.5469  &    0.0039  &     6.454  &    0.2782  &    -0.103  &    -4.183  &    0.0240  &    0.0079  &    0.0480  &      5.05  &      0.34 \\
               68\% CI (Upper)  &                           ...  &    0.0018  &    0.0007  &     0.032  &    0.0012  &     0.005  &     0.004  &    0.0002  &    0.0023  &    0.0008  &      0.04  &      0.04 \\
               68\% CI (Lower)  &                           ...  &   -0.0018  &   -0.0007  &    -0.032  &   -0.0012  &    -0.005  &    -0.004  &   -0.0002  &   -0.0030  &   -0.0007  &     -0.04  &     -0.04 \\
\hline
   Best-Fit &             6847.596  &    0.5469  &    0.0038  &     6.482  &    0.2787  &    -0.105  &    -4.182  &    0.0239  &    0.0091  &    0.0477  &      5.02  &      0.37 \\
\hline\hline
              $u_0 < 0$ Median  &                           ...  &    0.5474  &   -0.0039  &     6.455  &    0.2782  &    -0.104  &     4.183  &    0.0240  &    0.0038  &    0.0484  &      5.05  &      0.34 \\
               68\% CI (Upper)  &                           ...  &    0.0018  &    0.0007  &     0.032  &    0.0011  &     0.005  &     0.004  &    0.0002  &    0.0030  &    0.0006  &      0.04  &      0.03 \\
               68\% CI (Lower)  &                           ...  &   -0.0018  &   -0.0007  &    -0.031  &   -0.0011  &    -0.005  &    -0.004  &   -0.0002  &   -0.0023  &   -0.0006  &     -0.04  &     -0.04 \\
\hline
   Best-Fit  &             6847.872  &    0.5469  &   -0.0039  &     6.465  &    0.2782  &    -0.106  &     4.182  &    0.0239  &    0.0029  &    0.0482  &      5.03  &      0.36 \\
\hline\hline
\end{tabular}
\end{center}
Notes: \\
DoF = 7570$-$11 \\
\end{table*}

A comparison between the final fitted microlensing parameters and those from the heuristic assessment of \textsection \ref{ss:heuristic} reveals that they are qualitatively consistent. 

We also fit for higher order effects in the OGLE $I$-band light curve but we found that no meaningful constraints could be placed on the
annual parallax and orbital motion from these data alone. This is unsurprising because these phenomena typically manifest on timescales of tens to hundreds of days. In contrast, OB140962 is magnified for merely $\sim 10$ days, with only $\sim 2$ days between caustic crossings.

\bigskip
\section{Properties of the Source and Lens}\label{s:res}


\bigskip
\subsection{Source Star Limb Darkening and Angular Radius}\label{ss:sourcestar}

The source star properties can be inferred from its position on the local CMD. We calculate the apparent source ($V-I$) color by fitting a line to the observed event flux in the OGLE $V$ vs.\ OGLE $I$-band. This yields a model-independent color of $(V-I) = 1.963 \pm 0.006$. Modelling the source flux yields an apparent $I$-band magnitude of $16.22 \pm 0.01$. 

Using the observed ($V-I$) vs.\ $I$ CMD from OGLE in the field of the target, we locate the red clump. The apparent clump centroid is $(V-I, I)_{\rm clump} = (2.215, 15.59)$. According to \citet{Bensby13} and \citet{Nataf13}, the intrinsic color and magnitude of the red clump in the event direction is $(1.06, 14.36)$. Using the offset between the observed and actual clump centroid as a measure of source extinction and reddening \citep{Yoo04}, we determine the intrinsic source color and brightness to $(V-I, I)_{\rm 0, source} = (0.808, 14.98)$. From \citet{Bessell88}'’s color table for giants, we infer the source to be a G0 giant. 

We interpolate the color tables in \citet{Bessell88} to arrive at a $(V-K, K)_{\rm source} = (1.78 \pm 0.04, 14.00 \pm 0.06)$. Using the relationship between stellar angular size, ($V-K$) color, and $K$-magnitude given by \citet{Adams18}, the source angular radius $\theta_\star = 3.4 \pm 0.2 ~\mu {\rm as}$. 

The source star's brightness profile, which is important for modelling the caustic crossings, is parametrized by its limb darkening properties. \HL{\citet{Claret11} gives linear limb darkening coefficients ($u$) for stars with a variety of effective temperatures, surface gravities, metallicities, and microturbulences.} Assuming solar metallicity, we determine $T_{\rm eff}$ to be $5400 \pm 100$K using its relation with $V-I$ color from \citet{Casagrande10}. We adopt $\log(g) = 3$ and microturbulence $\sim 2$ km/s. \HL{The corresponding $\Gamma = (2u/(3-u))$ is $0.41$ in the $I$-band and $0.14$ in {\emph{Spitzer}} $L$-band.} These $\Gamma$ values are in turn used in the final fits of the light curve parameters as described in \textsection \ref{ss:modelling}.

\bigskip
\subsection{Physical Parameters of the Lens}\label{ss:phys-pars}

From the equations listed in \textsection \ref{ss:ulensing-pars}, the physical properties of the system are straightforwardly determined. We display the results in Table \ref{t:physprop}. The uncertainties are derived from direct propagation. We calculate $D_L$, $a_{\rm proj}$, and $D_{LS}$, assuming zero uncertainty in $D_S$. These final calculated properties are broadly compatible with the estimates from the earlier heuristic arguments. 

\begin{table}[htb]
\begin{center}
\caption{Binary Lens Physical Properties}
\label{t:physprop}
\begin{tabular}{lcc}
\hline\hline
Quantity & \multicolumn{2}{c}{Value from Best-Fit Solution} \\
 & $u_0 > 0$ & $u_0 < 0$ \\
\hline
                       $\pi_E$  &     0.049$\pm$     0.001  &     0.048$\pm$     0.001 \\
              $\theta_E$ (mas)  &     0.144$\pm$     0.008  &     0.144$\pm$     0.008 \\
      $\mu_{\rm rel}$ (mas/yr)  &     8.119$\pm$     0.001  &     8.132$\pm$     0.001 \\
        $M_{\rm tot} (M_\sun)$  &     0.365$\pm$     0.020  &     0.366$\pm$     0.020 \\
       $M_{\rm prim} (M_\sun)$  &     0.204$\pm$     0.011  &     0.205$\pm$     0.011 \\
        \HL{$M_{\rm sec} (M_\sun)$}  &     0.160$\pm$     0.009  &     0.161$\pm$     0.009 \\
$D_S$ (kpc)  &     7.863$\pm$     0.000  &     7.863$\pm$     0.000 \\
                   $D_L$ (kpc)  &     7.453$\pm$     0.021  &     7.455$\pm$     0.021 \\
           $a_{\rm proj}$ (AU)  &     2.040$\pm$     0.107  &     2.036$\pm$     0.107 \\
               $D_{8.3}$ (kpc)  &     7.845$\pm$     0.021  &     7.847$\pm$     0.021 \\
                      $D_{LS}$  &     0.410$\pm$     0.021  &     0.407$\pm$     0.021 \\
\hline\hline
\end{tabular}
\end{center}

\end{table}

Since the exact source distance is unknown, we list both $D_{\rm 8.3}$ \citep{CalchiNovati15a} and $D_L$, which is relative to the mean bar clump distance at the event's Galactic coordinates, which is $D_S = 7.86$ kpc \citep{Nataf13}. Regardless of the precise location of the source, this M+M-dwarf binary is the farthest lensing systems discovered with {\emph{Spitzer}}.  



\bigskip
\section{The {\emph{Spitzer}} Microlens Sample: a Testbed for Bayesian Analysis}\label{s:bayes}

OB140962 represents another addition to the growing sample of well-characterized microlensing systems from the {\emph{Spitzer}} satellite parallax campaign. This set of objects have directly measured $M_L$ and $D_L$, in contrast to most microlensing discoveries to date, whose physical properties are indeterminate due to the lack of constraint on $\pi_{\rm E}$. For the caustic-crossing microlensing events for which $t_{\rm E}$ and $\theta_{\rm E}$ are typically measured, the standard way to proceed is to perform a Bayesian analysis to infer probabilistic distributions for these parameters. This involves evaluating the likelihood of particular lens-source configurations using a Galactic model prior conditioned upon the measured $t_{\rm E}$ and $\theta_{\rm E}$ (taken together, they also encode the magnitude of the lens-source relative proper motion, see Equation (\ref{e:pie})). The resulting posteriors are often broad, with CIs spanning about a half-dex in mass and 2-3 kpc in distance. In Section \ref{ss:bayes-models}, we describe the ingredients that go into such an analysis. 

It is important to have confidence in the conclusions from the Bayesian analysis, since this will remain the chief channel for deriving lens system properties in the absence of expensive simultaneous satellite observations. They will continue to affect our understanding of both individual systems and ensemble statistics \citep[e.g.][]{CalchiNovati15a,Penny16}. 

The {\emph{Spitzer}} sample provides an excellent opportunity to test the accuracy of Bayesian analysis for $\sim$ a dozen systems (Table \ref{t:bayes1}). In Section \ref{ss:bayes-0962}, we demonstrate such a comparison for OB140962. Then, in Section \ref{ss:bayes-spitzer}, we extend this test to the subset of {\emph{Spitzer}} microlenses published to date with similarly secure characterizations.   



\bigskip
\subsection{Bayesian Formalism}\label{ss:bayes-models}

The Bayesian analysis framework is based on a Galactic model prior, whose ingredients are velocity distributions (VD), mass functions (MF), and density profiles (DP) of the bulge and disk of the Milky Way, in the direction of the event. Each draw of a lens-source pair has a corresponding $\theta_{\rm E}$ and $\mu_{\rm rel}$ (which is interchangeable with $t_{\rm E}$). For binary lenses, we make the following modification to the original Bayesian formalism, whose mass function assumes single stars and does not account for binaries. We draw the mass of the primary component, $M_{\rm prim}$, assuming that it follows the single-star MF from \citet{Chabrier03}. Then we calculate $M_{\rm tot}$ and $\theta_{\rm E}$ from $M_{\rm tot} = M_{\rm prim} (1+q)$. 
\HL{One underlying assumption is that the binary parameters of the event $(s, q)$ do not depend on the mass and distance of the lens.} The likelihood function is constrained by the observed $t_{\rm E}$ and $\theta_{\rm E}$ of the event. For a detailed description of the Galactic model used in this work, see \citet{Jung18}, which draws from \citet{Han95, Han03}, and \citet{Batista11}. Alternative models, varying some or all of the three components described above, are also used in literature \citep[e.g.][]{Bennett14, Zhu17a}.

\bigskip
\subsection{OB140962}\label{ss:bayes-0962}



\begin{figure}[htb]
\vspace{-0.5cm}
\hspace{-1cm}
\includegraphics[scale=0.55]{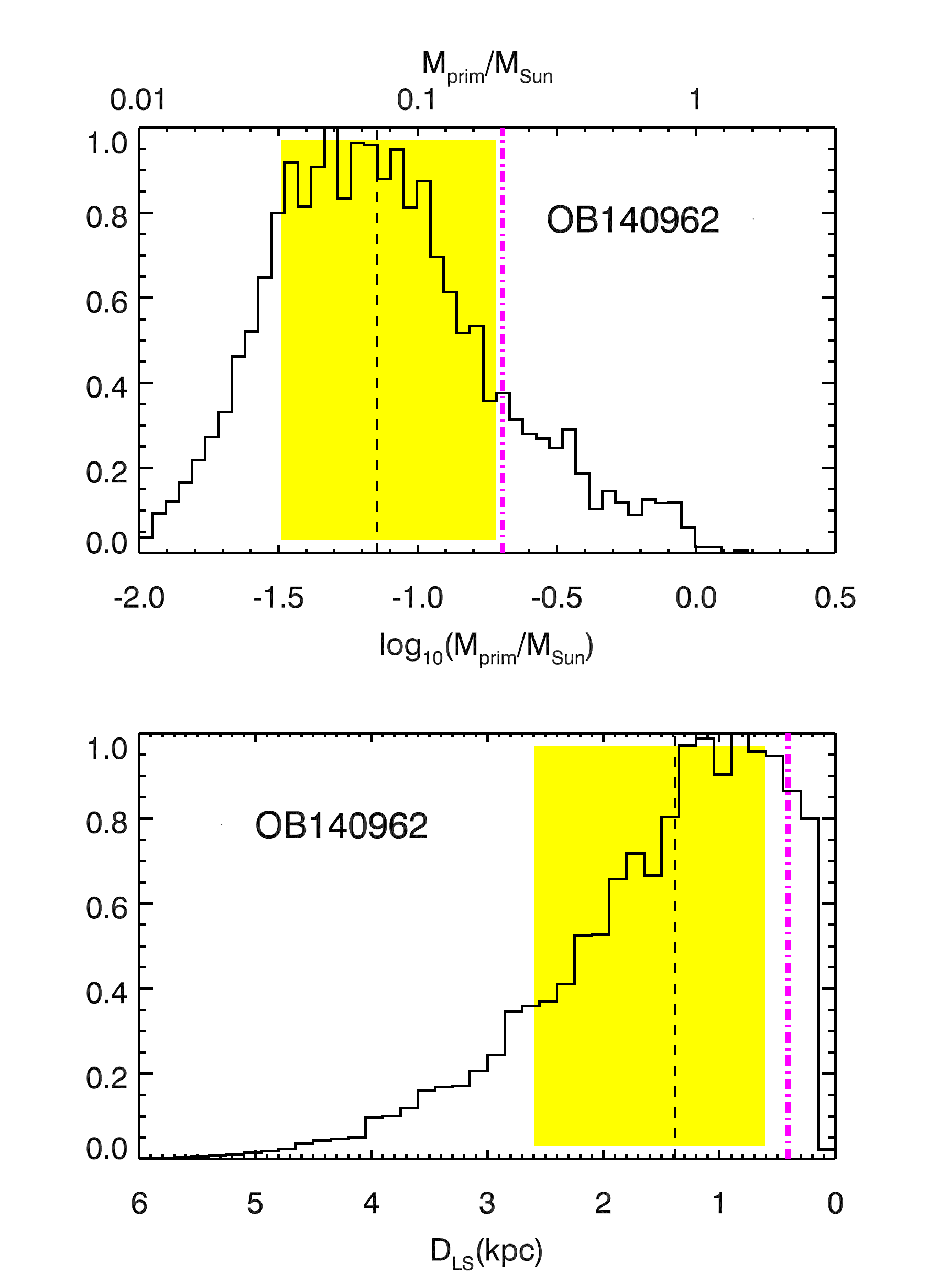}
\caption{Posteriors of lens physical properties ($M_{\rm prim}$ and $D_{\rm LS}$) for OB140962 based on Bayesian analysis. The Bayesian median is demarcated by the black dashed verticals, while the yellow shaded boxes outline the 68\% CIs about the median. Overplotted in the magenta dash-dotted lines are the values measured directly from {\emph{Spitzer}} parallax. Whereas the Bayesian analysis argues for a brown dwarf binary, the {\emph{Spitzer}} measurement clearly attributes the event to a mid-M+M-dwarf binary lens very close to the Galactic center. Only $\sim 14\%$ ($\sim 9\%$) of the Bayesian posterior lies above (below) the true $M_{\rm L}$ ($D_{\rm LS}$).  }
\vspace{+0.2cm}
\label{f:bayes-0962}
\end{figure}

We input our best-fit $\theta_{\rm E}$ and $t_{\rm E}$ values from the $u_0 > 0$ solution into the Bayesian analysis. Figure \ref{f:bayes-0962} displays the posterior distributions for $M_{\rm prim}$ and $D_{\rm LS}$. $D_{\rm LS}$ is a more robust metric than $D_{\rm L}$, since it is largely independent of the uncertainty in the source distance. The directly measured quantities are overplotted. Bayesian analysis indicates that the lens' primary has $\log (M_{\rm prim}/M_\sun) = -1.15^{+0.43}_{-0.35}$ or $M_{\rm prim} = 0.07^{+0.12}_{-0.039} M_\sun$. The true value of $M_{\rm prim} \sim 0.20$ is \HL{just outside} the 68\% CI. In this case, the difference between the Bayesian value and the true value means the difference between a brown-dwarf binary detection and a run-of-the-mill M+M binary. Similarly, the parallax-derived value of 0.41 kpc is outside the 68\% CI of the Bayesian prediction for the lens-source distance $D_{\rm LS} = 1.38^{+1.22}_{-0.77}$ kpc. 
The very small $\pi_{\rm E}$ measured by Spitzer places the true lens very close to the source star. It is clear that {\emph{Spitzer}} has provided important added value for accurately determining the mass and distance to this microlens. 

\bigskip
\subsection{Other {\emph{Spitzer}} Microlensing Systems}\label{ss:bayes-spitzer}

Bayesian analysis is expected to give a statistical representation of the truth. As such, it is not in itself surprising that individual outliers like OB140962 exist. The growing inventory of objects with {\emph{Spitzer}} satellite parallaxes allows us to investigate whether there are systematic problems with the Bayesian framework for a larger sample.

\setlength{\tabcolsep}{2.5pt}
\begin{table*}[htb]
\begin{center}
\caption{{\emph{Spitzer}} Bayesian \& Binary Sample Properties}
\label{t:bayes1}
\begin{tabular}{llrrcrrrccccccc}
\hline\hline
Object & Deg & $l$ & $b$ & $\theta_{\rm E}$ & $t_{\rm E}$ (days) & $\pi_E$ & $\mu_{{\rm hel}, l}$ & $\mu_{{\rm hel}, b}$ & $D_{\rm 8.3}$ & $M_{\rm L, tot}$ & $q$\textsuperscript{a} & Bayes\textsuperscript{b} & Bin\textsuperscript{c} &  Ref\textsuperscript{d} \\
Abbrv &Solution &  &  &  &  &  & (mas/yr) &(mas/yr) &(kpc) &($M_\sun$) & &  &  &  \\
\hline
            \HL{OB140124}  &        ++  &      2.34  &     -2.92  &     1.030  $\pm$     0.060  &    150.80  $\pm$      2.80  &     0.146  &      1.60  &     -3.02  &       3.7  &      0.90  &        6.9e-04  &    Y  &    Y  &   1, 15 \\
            OB140289  &        ++  &      0.80  &     -1.62  &     1.170  $\pm$     0.090  &    144.43  $\pm$      0.24  &     0.152  &      1.40  &      1.91  &       3.4  &      0.94  &        8.1e-01  &    Y  &    Y  &   2 \\
            OB140962  &            &      2.66  &     -2.54  &     0.144  $\pm$     0.008  &      6.45  $\pm$      0.03  &     0.049  &      5.12  &     -6.36  &       7.8  &      0.36  &        7.9e-01  &    Y  &    Y  &   3 \\
            OB141050  &        ++  &      5.09  &      3.23  &     1.340  $\pm$     0.160  &     73.30  $\pm$      4.20  &     0.120  &      1.48  &     -7.38  &       3.6  &      1.37  &        3.9e-01  &    Y  &    Y  &   4 \\
            OB150020  &            &     -2.24  &     -3.16  &     1.329  $\pm$     0.049  &     63.55  $\pm$      0.06  &     0.223  &     -5.88  &     -4.67  &       2.4  &      0.73  &        2.1e-01  &    Y  &    Y  &   5 \\
            OB150479  &        --  &     -5.82  &     -3.08  &     1.870  $\pm$     0.430  &     86.30  $\pm$      0.50  &     0.125  &     -6.09  &      2.86  &       2.8  &      1.83  &        8.1e-01  &    Y  &    Y  &   6 \\
            OB150763  &        -+  &     -1.85  &      2.25  &     0.288  $\pm$     0.020  &     32.78  $\pm$      0.25  &     0.071  &      2.94  &      1.32  &       7.1  &      0.50  &        \nodata  &    Y  &    N  &   7 \\
            OB150966  &  -- close  &      0.96  &     -1.82  &     0.760  $\pm$     0.070  &     57.80  $\pm$      0.40  &     0.241  &     -2.57  &      2.70  &       3.3  &      0.39  &        1.7e-04  &    Y  &    Y  &   8 \\
            OB151268  &            &      7.34  &      1.42  &     0.127  $\pm$     0.009  &     17.50  $\pm$      0.70  &     0.347  &      2.68  &     -0.97  &       6.1  &      0.05  &        \nodata  &    Y  &    N  &   7 \\
            OB151319  &   -+ wide  &     -1.71  &     -4.05  &     0.660  $\pm$     0.070  &     98.80  $\pm$      4.80  &     0.124  &     -0.02  &      2.00  &       4.9  &      0.65  &        9.5e-02  &    N  &    Y  &   9 \\
            OB160168  &            &     -1.84  &     -2.42  &     1.410  $\pm$     0.120  &     93.67  $\pm$      1.17  &     0.363  &      6.12  &      1.13  &       1.6  &      0.48  &        7.7e-01  &    Y  &    Y  &  10 \\
            OB161045  &        -+  &     -5.75  &     -1.39  &     0.245  $\pm$     0.015  &     11.98  $\pm$      0.08  &     0.355  &      5.86  &     -5.41  &       4.8  &      0.08  &        \nodata  &    Y  &    N  &  11 \\
            OB161190  &            &      2.62  &     -1.84  &     0.490  $\pm$     0.040  &     93.53  $\pm$      0.89  &     0.067  &      1.77  &      0.67  &       6.5  &      0.90  &        1.5e-02  &    Y  &    Y  &  12 \\
            OB161195  &  -+ close  &     -0.00  &     -2.48  &     0.286  $\pm$     0.050  &      9.96  $\pm$      0.11  &     0.473  &      0.29  &      9.76  &       3.9  &      0.07  &        5.5e-05  &    Y  &    Y  &  13 \\
            OB161266  &       A--  &     -0.04  &     -1.50  &     0.227  $\pm$     0.011  &      8.65  $\pm$      0.08  &     0.971  &      9.92  &     -0.17  &       2.9  &      0.03  &        7.6e-01  &    N  &    Y  &  14 \\
\hline\hline
\end{tabular}
\end{center}
Notes: \\
\textsuperscript{a}: \nodata denotes single lenses for which the notion of $q$ is not applicable  \\
\textsuperscript{b}: whether or not the object is included in the Bayesian analysis \\
\textsuperscript{c}: whether or not the object is included in the binarity analysis \\
\textsuperscript{d}: References: 1: \citet{Udalski15b}; 2: \citet{Udalski18}; 3: This Work; 4: \citet{Zhu15}; 5: \citet{Wang17}; 6: \citet{Han16}; 7: \citet{Zhu16}; 8: \citet{Street16}; 9: \citet{Shvartzvald16}; 10: \citet{Shin17}; 11: \citet{Shin18}; 12: \citet{Ryu18}; 13: \citet{Shvartzvald17}; 14: \citet{Albrow18}; 15: \citet{Beaulieu18} \\
\end{table*}

We repeat the Bayesian analysis for the 13 published {\emph{Spitzer}} events (including OB140962) with unique measurements of both $\pi_{\rm E}$ and $\theta_{\rm E}$. Their relevant properties are given in Table \ref{t:bayes1}. Among their ranks are both two-body lenses and single lenses with securely modelled finite-source effects. We treat binaries as described in \textsection \ref{ss:bayes-models}. For planetary events (defined for this purpose to be $q < 0.05$), we exclude stellar remnants from the Galactic models. For events with degeneracies for which the authors advocate strongly for one particular solution for $\theta_{\rm E}$ and $\pi_{\rm E}$ based either on $\chi^2$ fitness or on physical grounds (OB140289: \citealt{Udalski18}; OB141050: \citealt{Zhu15}; OB161045: \citealt{Shin18}; OB161190: \citealt{Ryu18}), we retain the favoured solutions only. Events with degeneracies yielding physical properties consistent within $1\sigma$ are assigned physical properties corresponding to the solution with the lowest $\chi^2$ for this comparison (OB140124: \citealt{Udalski15b, Beaulieu18}; OB141050: \citealt{Zhu15}; OB150479: \citealt{Han16}; OB161195: \citealt{Shvartzvald17}). \HL{Note that, although membership in our sample requires selection for {\emph{Spitzer}} followup, it is not necessarily true that both $\theta_E$ and $\pi_E$ were immediately constrained by the {\emph{Spitzer}} + ground observations. For OB140124, $\theta_{\rm E}$ is not well-measured due to the absence of the finite source effect. Therefore, we assign to it physical parameters determined by followup AO \citep{Beaulieu18}. In the case of OB140289, {\emph{Spitzer}} could not constrain $\pi_E$ because it happened to observe a featureless region of the light curve. Fortuitously, this event is sufficiently long such that annual parallax could be accurately and precisely determined. } 

We exclude from our Bayesian sample the events with severe degeneracies, i.e. those with multiple solutions for which the physical properties $M_L$ and $D_L$ are incompatible within their nominal uncertainties (OB150196: \citealt{Han17}; OB151482: \citealt{Chung17}; OB170329: \citealt{Han18}). \HL{It would be difficult to interpret a comparison between the Bayesian results and quantities that are ill-defined.} We also exclude OB151285 \citep{Shvartzvald15} and OB161266 \citep{Albrow18} from this exercise since the Bayesian priors for the mass function of planetary-mass objects and stellar remnants are not well understood. The literature does not homogeneously report the source distance assumed in the $D_L$ calculations. Therefore, for consistency, we calculate $D_{\rm LS}$ for each system using the clump distance in the direction of the event \citep{Nataf13} as $D_S$. 

\begin{figure}[htb]
\vspace{-1.5cm}
\hspace{-0.5cm}
\includegraphics[scale=0.6]{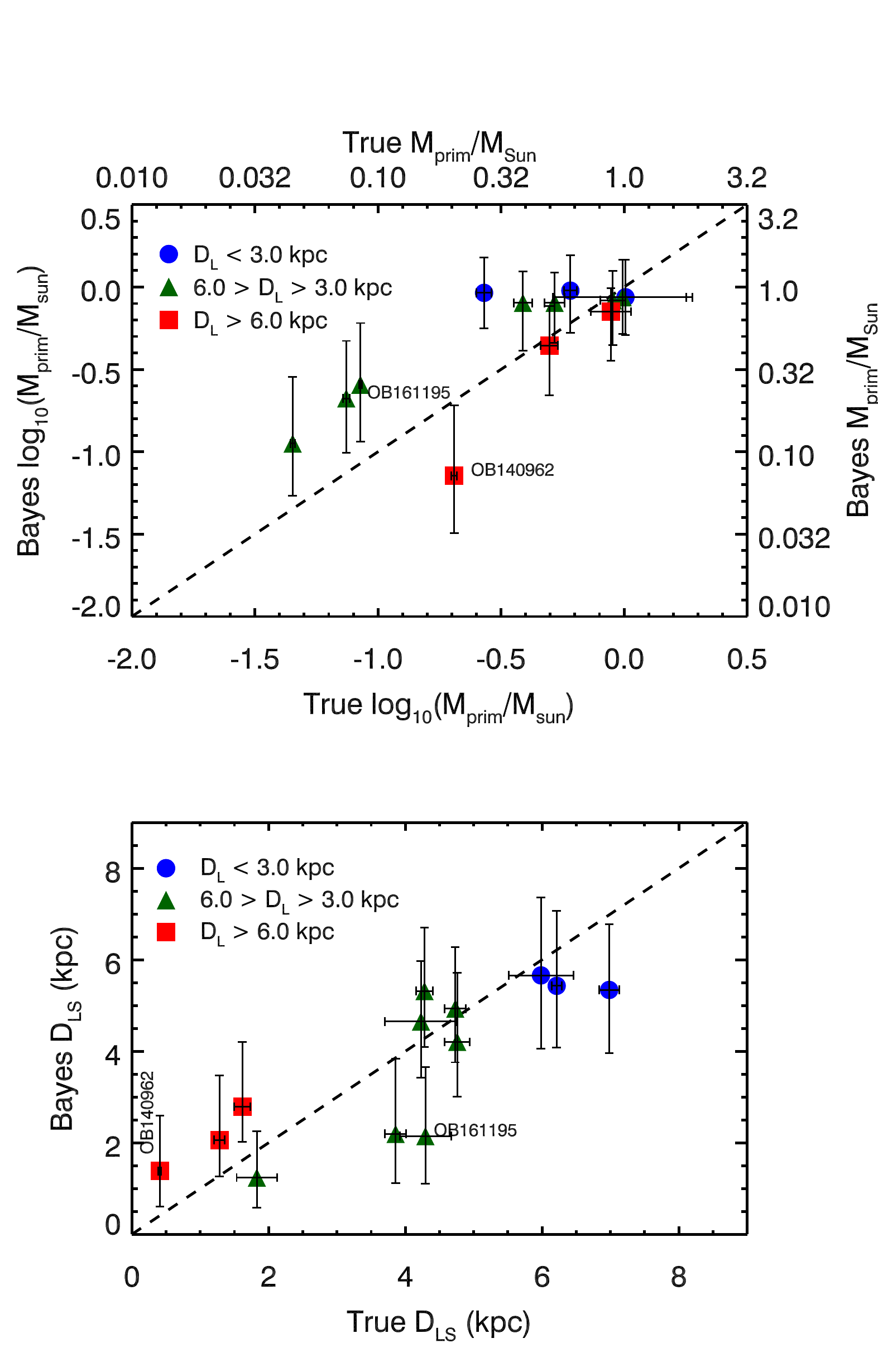}
\caption{Comparison between physical lens system properties $M_L$ and $D_{\rm LS}$ from Bayesian analysis and those derived from the parallax measurements, for {\emph{Spitzer}} lenses with unique, unambiguous solutions. \HL{For the most part, there is good agreement.} }
\vspace{+0.1cm}
\label{f:bayes-spitzer-1to1}
\end{figure}

 \setlength{\tabcolsep}{4pt}
\begin{table}[htb]
\begin{center}
\caption{\HL{{\emph{Spitzer}} Parallax-Bayesian Comparison}}
\label{t:bayes2}
\begin{tabular}{llrrrrrrrrrrrrll}
\hline\hline
Object & \multicolumn{2}{c}{Parallax} & \multicolumn{2}{c}{Bayesian\textsuperscript{a}} \\
Abbrv  & $M_{\rm prim} (M_\sun)$ & $D_{\rm LS}$ (kpc) & $M_{\rm prim} (M_\sun)$ & $D_{\rm LS}$ (kpc)\\
\hline
            OB140124  &      0.90  $\pm$      0.05  &      4.28  $\pm$      0.20  &    $      0.83  ^{+      0.43  }_{     -0.39   }$  &    $      5.32  ^{+      1.39  }_{     -1.22   }$ \\
            OB140289  &      0.52  $\pm$      0.04  &      4.73  $\pm$      0.16  &    $      0.80  ^{+      0.42  }_{     -0.34   }$  &    $      4.93  ^{+      1.34  }_{     -1.17   }$ \\
            OB140962  &      0.20  $\pm$      0.01  &      0.41  $\pm$      0.02  &    $      0.07  ^{+      0.12  }_{     -0.04   }$  &    $      1.38  ^{+      1.22  }_{     -0.77   }$ \\
            OB141050  &      0.99  $\pm$      0.28  &      4.23  $\pm$      0.57  &    $      0.87  ^{+      0.60  }_{     -0.35   }$  &    $      4.65  ^{+      1.32  }_{     -1.22   }$ \\
            OB150020  &      0.60  $\pm$      0.03  &      6.21  $\pm$      0.07  &    $      0.95  ^{+      0.60  }_{     -0.42   }$  &    $      5.43  ^{+      1.64  }_{     -1.35   }$ \\
            OB150479  &      1.01  $\pm$      0.25  &      5.98  $\pm$      0.45  &    $      0.87  ^{+      0.60  }_{     -0.36   }$  &    $      5.66  ^{+      1.70  }_{     -1.60   }$ \\
            OB150763  &      0.50  $\pm$      0.03  &      1.28  $\pm$      0.07  &    $      0.44  ^{+      0.33  }_{     -0.22   }$  &    $      2.06  ^{+      1.41  }_{     -0.80   }$ \\
            OB150966  &      0.39  $\pm$      0.04  &      4.76  $\pm$      0.19  &    $      0.80  ^{+      0.44  }_{     -0.39   }$  &    $      4.21  ^{+      1.51  }_{     -1.20   }$ \\
            OB151268  &      0.05  $\pm$      0.01  &      1.83  $\pm$      0.35  &    $      0.11  ^{+      0.17  }_{     -0.06   }$  &    $      1.24  ^{+      1.01  }_{     -0.66   }$ \\
            OB160168  &      0.27  $\pm$      0.03  &      6.98  $\pm$      0.14  &    $      0.92  ^{+      0.59  }_{     -0.36   }$  &    $      5.34  ^{+      1.43  }_{     -1.38   }$ \\
            OB161045  &      0.08  $\pm$      0.01  &      3.86  $\pm$      0.15  &    $      0.26  ^{+      0.35  }_{     -0.14   }$  &    $      2.19  ^{+      1.65  }_{     -1.08   }$ \\
            OB161190  &      0.88  $\pm$      0.08  &      1.62  $\pm$      0.13  &    $      0.71  ^{+      0.28  }_{     -0.35   }$  &    $      2.79  ^{+      1.42  }_{     -0.77   }$ \\
            OB161195  &      0.07  $\pm$      0.01  &      4.29  $\pm$      0.38  &    $      0.21  ^{+      0.26  }_{     -0.11   }$  &    $      2.15  ^{+      1.51  }_{     -1.04   }$ \\
\hline\hline
\end{tabular}
\end{center}
Notes: \\
\textsuperscript{a}: Bayesian values shown are median and symmetric 68\% CIs. \\

\end{table}

The result is summarized in Figure \ref{f:bayes-spitzer-1to1}. \HL{For the majority of the cases, the Bayesian posteriors are consistent with the true measured values. One of the outliers in the Bayesian prediction compared to the result derived from parallax belongs to OB140962} (see Section \ref{ss:bayes-0962}), \HL{for which the true value lies above the 84th percentile of its Bayesian posterior. However, if the posteriors are true representations of the data, occasional outliers are expected. In fact, this particular situation should occur for 16\% of the instances. } 

\begin{figure}[htb]
\vspace{-0.5cm}
\hspace{-1.5cm}
\includegraphics[scale=0.6]{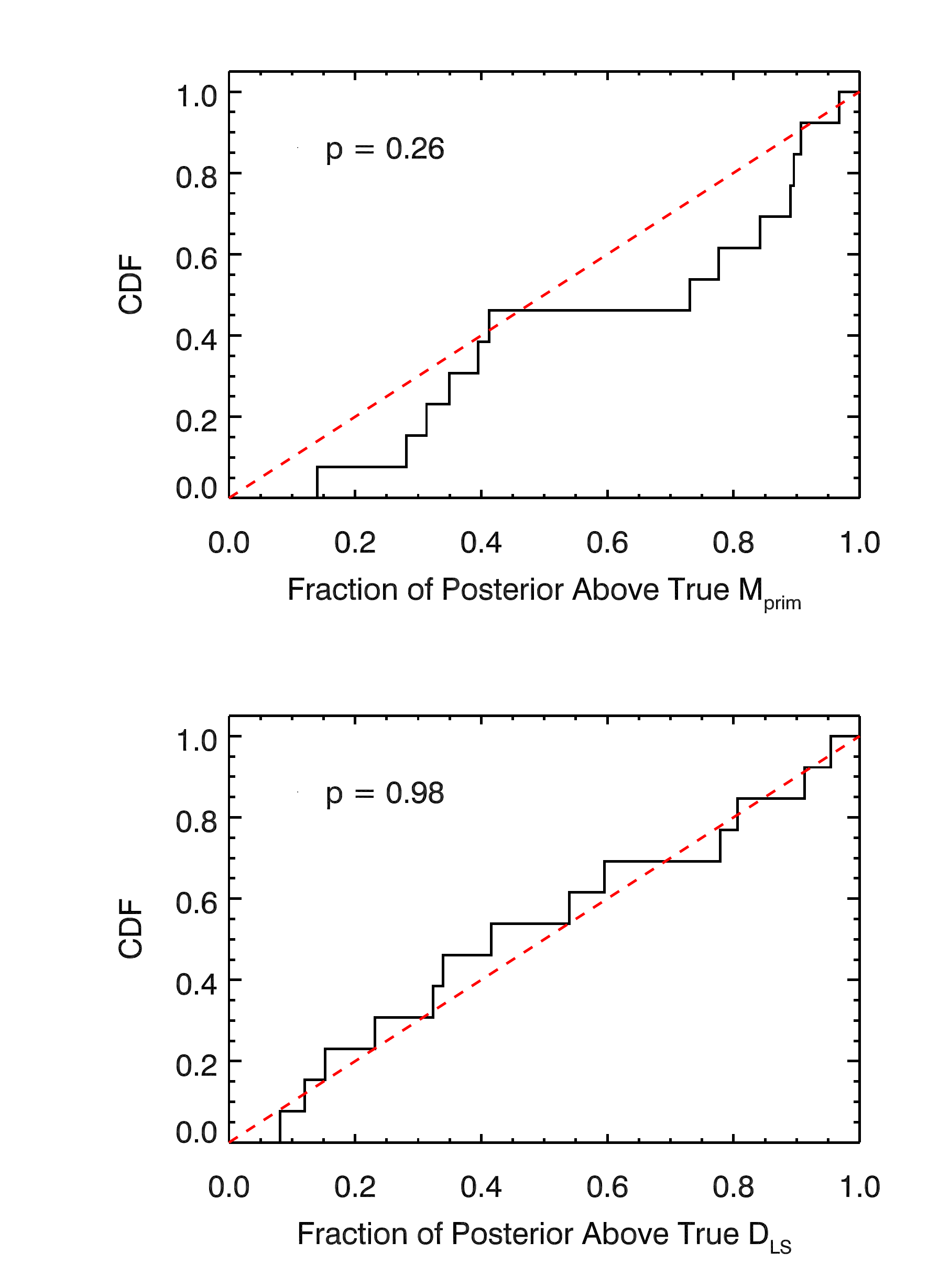}
\caption{Empirical cumulative distribution function of Bayesian posterior fraction above the lens system's physical properties ($M_L$ and $D_{\rm LS}$) derived from parallax, for {\emph{Spitzer}} lenses with unique, unambiguous solutions. The distributions are consistent with 1-to-1, suggesting that the Bayesian posteriors are a fair representation of the underlying true parameters. }
\vspace{+0.1cm}
\label{f:bayes-spitzer-cdf}
\end{figure}

\HL{To test the overall consistency of the Bayesian analysis, we evaluate the fraction of posterior lying above the true values of $M_L$ and $D_{\rm LS}$ derived from parallax for each object in this ensemble. If these Bayesian posteriors represent the true values fairly for this ensemble, we should expect this CDF to follow the identity function (e.g. 10\% of the time the true value falls below 10\% of the posterior). In Figure \ref{f:bayes-spitzer-cdf} we show the cumulative distribution of the fraction of posterior lying above the parallax value for $M_L$ and $D_{\rm LS}$. One-sample Kolmogrov-Smirnov (KS) tests show that neither parameters are significantly differently distributed from the 1-to-1 line. This indicates that, on average, Bayesian analysis is a good reflection of the data.} 

We note that, for OB161195, the physical properties derived from parallax are $M_L = 0.078^{+0.016}_{-0.012} M_\sun$ and $D_L = 3.91^{+0.42}_{-0.46}$ kpc $\rightarrow D_{\rm LS} \sim 4.29$ kpc, \citep{Shvartzvald17}. \HL{This is between the 68\% and 95\% CI of both our Bayesian posterior} ($M_L = 0.21^{+0.26}_{-0.11} M_\sun$ and $D_{\rm LS} = 2.15^{+1.51}_{-1.04}$ kpc), which is based on the $t_{\rm E}$ and $\theta_{\rm E}$ given in \citet{Shvartzvald17}, as well as that of \citet{Bond17}, which yielded $M_L = 0.37^{+0.38}_{-0.21}$ and $D_L = 7.20^{+0.85}_{-1.02}$. The discrepancy between our work and that of \citet{Bond17} is likely due primarily to differences in the Galactic models assumed. An important caveat is that OB161195 is kinematically peculiar: despite its disklike distance, its motion is not in the direction of the disk's rotation \citep{Shvartzvald17}. For a Bayesian analysis based on $\theta_E$ and $t_E$, the direction of the motion would be unknown. However, with this knowledge we recognize that a Bayesian analysis may not accurately reflect the properties of OB161195.  

\bigskip
\section{The Galactic Distribution of {\emph{Spitzer}} Binary Lenses} \label{s:binarity}

Ultimately, the {\emph{Spitzer}} planetary systems will be analyzed to determine whether or not planet occurrence varies across the Galaxy. One way to do this is to compare the distance distribution of planetary lenses to that of the {\emph{Spitzer}} single-lens sample \citep{CalchiNovati15a,Zhu17a}. Note that \citet{Penny16} undertook a related study, in which they compared ground-based planet discoveries with a simulated host population. 

While the {\emph{Spitzer}} planet sample is not yet large enough to perform this test, the number of {\emph{Spitzer}} 2-body lenses (both planets and binaries) is now comparable to the total number of planets expected for the full {\emph{Spitzer}} sample. Thus, we can use the {\emph{Spitzer}} binaries to illustrate the methodology for comparing distance distributions. To date, 12 binary lenses from {\emph{Spitzer}} have published parameters with unambiguous or strongly preferred solutions, and therefore $D_{8.3}$ (Table \ref{t:bayes1}). In this count we excluded OB151212 \citep{Bozza16} because only 3/8 degenerate solutions have constrained $D_{8.3}$. We also discarded OB150196 \citep{Han17} and OB170329 \citep{Han18} because they have severe discrete degeneracies. 

\begin{figure}[t!]
\vspace{-0.5cm}
\hspace{-1.5cm}
\includegraphics[scale=0.6]{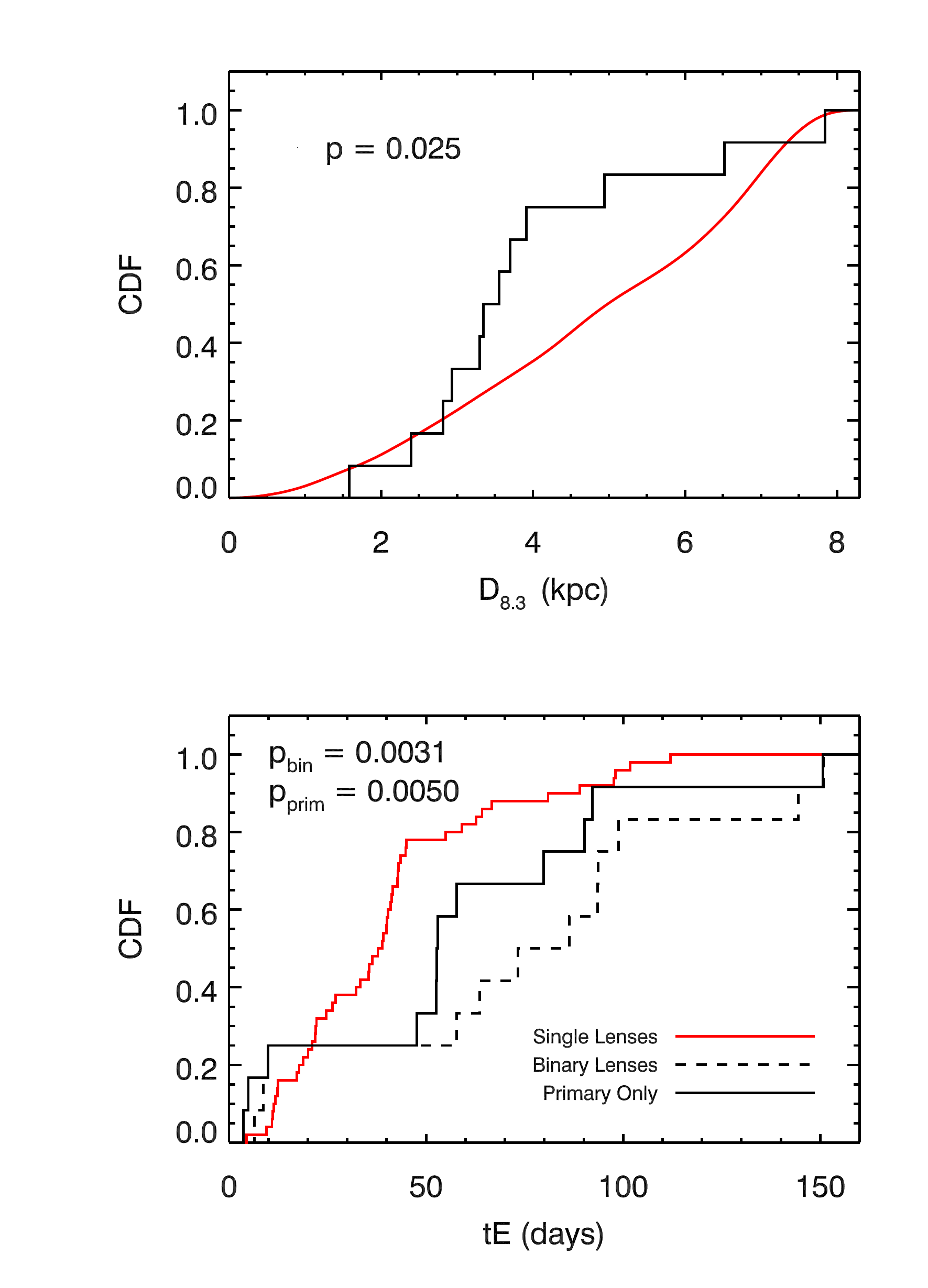}
\caption{Top Panel: The empirical cumulative distance distribution function of {\emph{Spitzer}} binary lenses (black), overplotted on the {\emph{Spitzer}} single lens detections from 2015 (red). The binary ensemble has only a \HL{2.5\%} probability of being drawn from the single lens distribution. Bottom Panel: The cumulative distribution of $t_E$ for single lenses (red) and binary lenses (black), with the distribution of $t_E$ calculated from the primary component of the binary lens denoted by the black solid line. Based on a 2-sample KS test, the single and binary lens distributions are discrepant at the $3\sigma$ level.}
\label{f:binary-cdf}
\end{figure}

Figure \ref{f:binary-cdf} shows the empirical binary cumulative distance distribution function for these 12 events. Note that, although OB151319 \citep{Shvartzvald16} was excluded from the Bayesian exercise because its 8 degenerate solutions span primary masses of 0.53 to 0.67 $M_\sun$ and fail our criteria for mass consistency, it is included here as one entry (with $D_{8.3}$ of the solution with the best $\chi^2$) because the $D_{8.3}$ for all degenerate solutions are actually all consistent with each other. The substellar binary candidate OB161266 \citep{Albrow18}, previously excluded from the Bayesian sample (see Section \ref{ss:bayes-spitzer}), also enters into this analysis using $D_{8.3}$ from the `A' solution.   

Overplotted on Figure \ref{f:binary-cdf} is the cumulative distance distribution of {\emph{Spitzer}} single lens detections from \citet{Zhu17a} (the `Standard' distribution from Figure 12). An excess of binary lenses is visually apparent at intermediate distances ($D_{8.3} =$ 3 to 5 kpc). 
To quantify this visual discrepancy, we perform a 1-sample KS-test for the empirical CDF. Formally, \HL{at a $p$-value of 0.025}, the null hypothesis that the binary and single lens sample are drawn from the same distribution can be rejected at $2\sigma$ significance. 

Investigating the source of this intermediate-distance excess relative to the bulge is outside the scope of this paper, since any physical conclusions would first require disentangling the contribution from selection effects. While the selection effects should be the same for planets compared to single lenses, this is not \HL{necessarily} true for binaries \citep[cf.][]{Yee15a}. \HL{For example, some binaries are discovered serendipitously as part of the single-lens sample selected for {\emph{Spitzer}} followup, whereas others are deliberately observed by {\emph{Spitzer}} after their binary anomalies are already detected from the ground. One indication of possible selection bias between single and 2-body {\emph{Spitzer}} lenses is displayed in the bottom panel of Figure \ref{f:binary-cdf}, which shows a $3\sigma$ discrepancy (from a 2-sample KS test) between the $t_E$ distributions of single and binary lenses.} The Figure gives the $t_E$ directly from the model (which is relative to the total mass of the system) and the $t_E$ relative to the primary alone. The latter is the most relevant comparison for single lesnses because it shows what would have been observed in the absence of a companion. The discrepancy in the $t_E$ distribution likely contributes to the excess at intermediate distances, but even the origin of the $t_E$ discrepancy is unknown (possibly related to selection effects). Regardless of the physical explanation for the excess at 3 -- 5 kpc, the above analysis shows that this method can measure significant differences in the distance distribution of a population with only 12 objects relative to the single-lens population. 

\bigskip
\section{Summary} \label{s:sum}

Measurements of the Einstein radius ($\theta_{\rm E}$) and the microlens parallax ($\pi_{\rm E}$) make a powerful combination for deducing the physical properties of lensing systems, such as mass, distance, and kinematics. This information is readily available for events involving binary lenses with satellite observations. However, for many ground-based microlensing discoveries it is not possible to obtain both quantities for an unambiguous solution. In these cases, a Bayesian analysis, based on a Galactic model, is used to give a probabilistic estimate of the physical parameters. The interpretation of many individual systems and ensemble statistics depends on the accuracy of the Bayesian analysis, which we systematically verify in this work.    

We first present the discovery and characterization of OGLE-2014-BLG-0962 using high quality ground-based survey and {\emph{Spitzer}} data. The densely covered light curves allow us to constrain $\theta_{\rm E}$ and $\vec{\pi}_{\rm E}$ very well, leading to a unique interpretation of this object to be a textbook mid-M-M stellar binary deeply embedded in the Galactic bulge. However, the angular Einstein radius (0.14 mas) is on the low side. Thus, if we were to infer the physical properties of this system without using the parallax information -- that is, using a standard Bayesian analysis based on a Galactic model prior -- we would infer a much lower lens mass.    

To investigate whether the Bayesian framework is on average reliable, we assemble a sample of 13 well-understood {\emph{Spitzer}} systems and perform Bayesian analysis on each of them using their $t_{\rm E}$ and $\theta_{\rm E}$ as inputs. We compare the Bayesian predictions of lens mass ($M_L$) and lens-source distance ($D_{\rm LS}$) to the same physical properties calculated from $\pi_{\rm E}$, finding good agreement overall and concluding that the Bayesian posteriors are on average representative of the true answers. 

We also construct a sample of {\emph{Spitzer}} binaries and show the methodology for making quantitative statements about the Galactic distribution of planetary and binary lenses using detections from {\emph{Spitzer}}. A comparison with that of single lens detections from {\emph{Spitzer}} shows tentative evidence that the two types of lenses are drawn from incompatible distance distributions. Specifically, binaries may be more abundant relative to single stars at the intermediate distances (i.e. 3-5 kpc) and deficient beyond $\sim 6$ kpc, the latter coinciding with the geographical location of the Galactic bulge. We do not investigate the reason for this discrepancy, which could be related to the difference in the $t_E$ distributions or other selection effects. However, while understanding the exact source of this excess lies outside the scope of this work, we have demonstrated our ability to measure a significant difference between a reference spatial distribution function and that of a sample of interest with just 12 objects. Our finding bodes well for the primary mission of the {\emph{Spitzer}} microlensing campaign to constrain the Galactic distribution of planets using a similarly sized sample.  


\bigskip
\section*{Acknowledgments}
The authors would like to thank John Johnson, Jason Eastman, and other members of the Exolab for providing helpful suggestions and feedback throughout the course of this work. We are grateful to Vinay Kashyap for his statistical guidance. We thank Wei Zhu for contributing the data for the {\emph{Spitzer}} single lens CDF (i.e., Figure 12 of \citet{Zhu17a}) and for discussions. Y. Shan is supported in part by a Doctoral Postgraduate Scholarship from the Natural Science and Engineering Research Council (NSERC) of Canada. The OGLE project has received funding from the National Science Centre,
Poland, grant MAESTRO 2014/14/A/ST9/00121 to AU. OGLE Team thanks Profs.
M.~Kubiak and G.~Pietrzy{\'n}ski, for their contribution to the OGLE
photometric dataset. The MOA project is supported by JSPS KAKENHI Grant Number JSPS24253004, JSPS26247023, JSPS23340064, JSPS15H00781, and JP16H06287. This research was supported by the I-CORE programme
of the Planning and Budgeting Committee and the Israel
Science Foundation, Grant 1829/12. DM and AG acknowledge support by
the US-Israel Binational Science Foundation.





\begin{thebibliography}{}
\expandafter\ifx\csname natexlab\endcsname\relax\def\natexlab#1{#1}\fi
\providecommand{\url}[1]{\href{#1}{#1}}
\providecommand{\dodoi}[1]{doi:~\href{http://doi.org/#1}{\nolinkurl{#1}}}
\providecommand{\doeprint}[1]{\href{http://ascl.net/#1}{\nolinkurl{http://ascl.net/#1}}}
\providecommand{\doarXiv}[1]{\href{https://arxiv.org/abs/#1}{\nolinkurl{https://arxiv.org/abs/#1}}}

\bibitem[{{Adams} {et~al.}(2018){Adams}, {Boyajian}, \& {von Braun}}]{Adams18}
{Adams}, A.~D., {Boyajian}, T.~S., \& {von Braun}, K. 2018, \mnras, 473, 3608,
  \dodoi{10.1093/mnras/stx2367}

\bibitem[{{Alard} \& {Lupton}(1998)}]{Alard98}
{Alard}, C., \& {Lupton}, R.~H. 1998, \apj, 503, 325, \dodoi{10.1086/305984}

\bibitem[{{Albrow} {et~al.}(2009){Albrow}, {Horne}, {Bramich}, {Fouqu{\'e}},
  {Miller}, {Beaulieu}, {Coutures}, {Menzies}, {Williams}, {Batista},
  {Bennett}, {Brillant}, {Cassan}, {Dieters}, {Dominis Prester}, {Donatowicz},
  {Greenhill}, {Kains}, {Kane}, {Kubas}, {Marquette}, {Pollard}, {Sahu},
  {Tsapras}, {Wambsganss}, \& {Zub}}]{Albrow09}
{Albrow}, M.~D., {Horne}, K., {Bramich}, D.~M., {et~al.} 2009, \mnras, 397,
  2099, \dodoi{10.1111/j.1365-2966.2009.15098.x}

\bibitem[{{Albrow} {et~al.}(2018){Albrow}, {Yee}, {Udalski}, {Calchi Novati},
  {Carey}, {Henderson}, {Beichman}, {Bryden}, {Gaudi}, {Shvartzvald},
  {Szyman{\'s}ki}, {Mro{\'z}}, {Skowron}, {Poleski}, {Soszyn{\'s}ki},
  {Kozlowski}, {Pietrukowicz}, {Ulaczyk}, {Pawlak}, {Chung}, {Gould}, {Han},
  {Hwang}, {Jung}, {Ryu}, {Shin}, {Zhu}, {Cha}, {Kim}, {Kim}, {Kim}, {Lee},
  {Lee}, {Lee}, {Park}, \& {Pogge}}]{Albrow18}
{Albrow}, M.~D., {Yee}, J.~C., {Udalski}, A., {et~al.} 2018, ArXiv e-prints.
\newblock \doarXiv{1802.09563}

\bibitem[{{Batista} {et~al.}(2015){Batista}, {Beaulieu}, {Bennett}, {Gould},
  {Marquette}, {Fukui}, \& {Bhattacharya}}]{Batista15}
{Batista}, V., {Beaulieu}, J.-P., {Bennett}, D.~P., {et~al.} 2015, \apj, 808,
  170, \dodoi{10.1088/0004-637X/808/2/170}

\bibitem[{{Batista} {et~al.}(2011){Batista}, {Gould}, {Dieters}, {Dong},
  {Bond}, {Beaulieu}, {Maoz}, {Monard}, {Christie}, {McCormick}, {Albrow},
  {Horne}, {Tsapras}, {Burgdorf}, {Calchi Novati}, {Skottfelt}, {Caldwell},
  {Koz{\l}owski}, {Kubas}, {Gaudi}, {Han}, {Bennett}, {An}, {MOA
  Collaboration}, {Abe}, {Botzler}, {Douchin}, {Freeman}, {Fukui}, {Furusawa},
  {Hearnshaw}, {Hosaka}, {Itow}, {Kamiya}, {Kilmartin}, {Korpela}, {Lin},
  {Ling}, {Makita}, {Masuda}, {Matsubara}, {Miyake}, {Muraki}, {Nagaya},
  {Nishimoto}, {Ohnishi}, {Okumura}, {Perrott}, {Rattenbury}, {Saito},
  {Sullivan}, {Sumi}, {Sweatman}, {Tristram}, {von Seggern}, {Yock}, {PLANET
  Collaboration}, {Brillant}, {Calitz}, {Cassan}, {Cole}, {Cook}, {Coutures},
  {Dominis Prester}, {Donatowicz}, {Greenhill}, {Hoffman}, {Jablonski}, {Kane},
  {Kains}, {Marquette}, {Martin}, {Martioli}, {Meintjes}, {Menzies},
  {Pedretti}, {Pollard}, {Sahu}, {Vinter}, {Wambsganss}, {Watson}, {Williams},
  {Zub}, {FUN Collaboration}, {Allen}, {Bolt}, {Bos}, {DePoy}, {Drummond},
  {Eastman}, {Gal-Yam}, {Gorbikov}, {Higgins}, {Janczak}, {Kaspi}, {Lee},
  {Mallia}, {Maury}, {Monard}, {Moorhouse}, {Morgan}, {Natusch}, {Ofek},
  {Park}, {Pogge}, {Polishook}, {Santallo}, {Shporer}, {Spector}, {Thornley},
  {Yee}, {MiNDSTEp Consortium}, {Bozza}, {Browne}, {Dominik}, {Dreizler},
  {Finet}, {Glitrup}, {Grundahl}, {Harps{\o}e}, {Hessman}, {Hinse},
  {Hundertmark}, {J{\o}rgensen}, {Liebig}, {Maier}, {Mancini}, {Mathiasen},
  {Rahvar}, {Ricci}, {Scarpetta}, {Southworth}, {Surdej}, {Zimmer}, {RoboNet
  Collaboration}, {Allan}, {Bramich}, {Snodgrass}, {Steele}, \&
  {Street}}]{Batista11}
{Batista}, V., {Gould}, A., {Dieters}, S., {et~al.} 2011, \aap, 529, A102,
  \dodoi{10.1051/0004-6361/201016111}

\bibitem[{{Batista} {et~al.}(2014){Batista}, {Beaulieu}, {Gould}, {Bennett},
  {Yee}, {Fukui}, {Gaudi}, {Sumi}, \& {Udalski}}]{Batista14}
{Batista}, V., {Beaulieu}, J.-P., {Gould}, A., {et~al.} 2014, \apj, 780, 54,
  \dodoi{10.1088/0004-637X/780/1/54}

\bibitem[{{Beaulieu} {et~al.}(2018){Beaulieu}, {Batista}, {Bennett},
  {Marquette}, {Blackman}, {Cole}, {Coutures}, {Danielski}, {Dominis Prester},
  {Donatowicz}, {Fukui}, {Koshimoto}, {Lon{\v c}ari{\'c}}, {Morales}, {Sumi},
  {Suzuki}, {Henderson}, {Shvartzvald}, \& {Beichman}}]{Beaulieu18}
{Beaulieu}, J.-P., {Batista}, V., {Bennett}, D.~P., {et~al.} 2018, \aj, 155,
  78, \dodoi{10.3847/1538-3881/aaa293}

\bibitem[{{Bennett} {et~al.}(2014){Bennett}, {Batista}, {Bond}, {Bennett},
  {Suzuki}, {Beaulieu}, {Udalski}, {Donatowicz}, {Bozza}, {Abe}, {Botzler},
  {Freeman}, {Fukunaga}, {Fukui}, {Itow}, {Koshimoto}, {Ling}, {Masuda},
  {Matsubara}, {Muraki}, {Namba}, {Ohnishi}, {Rattenbury}, {Saito}, {Sullivan},
  {Sumi}, {Sweatman}, {Tristram}, {Tsurumi}, {Wada}, {Yock}, {MOA
  Collaboration}, {Albrow}, {Bachelet}, {Brillant}, {Caldwell}, {Cassan},
  {Cole}, {Corrales}, {Coutures}, {Dieters}, {Dominis Prester}, {Fouqu{\'e}},
  {Greenhill}, {Horne}, {Koo}, {Kubas}, {Marquette}, {Martin}, {Menzies},
  {Sahu}, {Wambsganss}, {Williams}, {Zub}, {PLANET Collaboration}, {Choi},
  {DePoy}, {Dong}, {Gaudi}, {Gould}, {Han}, {Henderson}, {McGregor}, {Lee},
  {Pogge}, {Shin}, {Yee}, {{$\mu$}FUN Collaboration}, {Szyma{\'n}ski},
  {Skowron}, {Poleski}, {Koz{\l}owski}, {Wyrzykowski}, {Kubiak},
  {Pietrukowicz}, {Pietrzy{\'n}ski}, {Soszy{\'n}ski}, {Ulaczyk}, {OGLE
  Collaboration}, {Tsapras}, {Street}, {Dominik}, {Bramich}, {Browne},
  {Hundertmark}, {Kains}, {Snodgrass}, {Steele}, {RoboNet Collaboration},
  {Dekany}, {Gonzalez}, {Heyrovsk{\'y}}, {Kandori}, {Kerins}, {Lucas},
  {Minniti}, {Nagayama}, {Rejkuba}, {Robin}, \& {Saito}}]{Bennett14}
{Bennett}, D.~P., {Batista}, V., {Bond}, I.~A., {et~al.} 2014, \apj, 785, 155,
  \dodoi{10.1088/0004-637X/785/2/155}

\bibitem[{{Bennett} {et~al.}(2015){Bennett}, {Bhattacharya}, {Anderson},
  {Bond}, {Anderson}, {Barry}, {Batista}, {Beaulieu}, {DePoy}, {Dong}, {Gaudi},
  {Gilbert}, {Gould}, {Pfeifle}, {Pogge}, {Suzuki}, {Terry}, \&
  {Udalski}}]{Bennett15}
{Bennett}, D.~P., {Bhattacharya}, A., {Anderson}, J., {et~al.} 2015, \apj, 808,
  169, \dodoi{10.1088/0004-637X/808/2/169}

\bibitem[{{Bensby} {et~al.}(2013){Bensby}, {Yee}, {Feltzing}, {Johnson},
  {Gould}, {Cohen}, {Asplund}, {Mel{\'e}ndez}, {Lucatello}, {Han}, {Thompson},
  {Gal-Yam}, {Udalski}, {Bennett}, {Bond}, {Kohei}, {Sumi}, {Suzuki}, {Suzuki},
  {Takino}, {Tristram}, {Yamai}, \& {Yonehara}}]{Bensby13}
{Bensby}, T., {Yee}, J.~C., {Feltzing}, S., {et~al.} 2013, \aap, 549, A147,
  \dodoi{10.1051/0004-6361/201220678}

\bibitem[{{Bessell} \& {Brett}(1988)}]{Bessell88}
{Bessell}, M.~S., \& {Brett}, J.~M. 1988, \pasp, 100, 1134,
  \dodoi{10.1086/132281}

\bibitem[{{Bond} {et~al.}(2001){Bond}, {Abe}, {Dodd}, {Hearnshaw}, {Honda},
  {Jugaku}, {Kilmartin}, {Marles}, {Masuda}, {Matsubara}, {Muraki}, {Nakamura},
  {Nankivell}, {Noda}, {Noguchi}, {Ohnishi}, {Rattenbury}, {Reid}, {Saito},
  {Sato}, {Sekiguchi}, {Skuljan}, {Sullivan}, {Sumi}, {Takeuti}, {Watase},
  {Wilkinson}, {Yamada}, {Yanagisawa}, \& {Yock}}]{Bond01}
{Bond}, I.~A., {Abe}, F., {Dodd}, R.~J., {et~al.} 2001, \mnras, 327, 868,
  \dodoi{10.1046/j.1365-8711.2001.04776.x}

\bibitem[{{Bond} {et~al.}(2017){Bond}, {Bennett}, {Sumi}, {Udalski}, {Suzuki},
  {Rattenbury}, {Bozza}, {Koshimoto}, {Abe}, {Asakura}, {Barry},
  {Bhattacharya}, {Donachie}, {Evans}, {Fukui}, {Hirao}, {Itow}, {Li}, {Ling},
  {Masuda}, {Matsubara}, {Muraki}, {Nagakane}, {Ohnishi}, {Ranc}, {Saito},
  {Sharan}, {Sullivan}, {Tristram}, {Yamada}, {Yamada}, {Yonehara}, {Skowron},
  {Szyma{\'n}ski}, {Poleski}, {Mr{\'o}z}, {Soszy{\'n}ski}, {Pietrukowicz},
  {Koz{\l}owski}, {Ulaczyk}, \& {Pawlak}}]{Bond17}
{Bond}, I.~A., {Bennett}, D.~P., {Sumi}, T., {et~al.} 2017, \mnras, 469, 2434,
  \dodoi{10.1093/mnras/stx1049}

\bibitem[{{Bozza} {et~al.}(2016){Bozza}, {Shvartzvald}, {Udalski}, {Calchi
  Novati}, {Bond}, {Han}, {Hundertmark}, {Poleski}, {Pawlak}, {Szyma{\'n}ski},
  {Skowron}, {Mr{\'o}z}, {Koz{\l}owski}, {Wyrzykowski}, {Pietrukowicz},
  {Soszy{\'n}ski}, {Ulaczyk}, {OGLE Group}, {and}, {Beichman}, {Bryden},
  {Carey}, {Fausnaugh}, {Gaudi}, {Gould}, {Henderson}, {Pogge}, {Wibking},
  {Yee}, {Zhu}, {Spitzer Team}, {Abe}, {Asakura}, {Barry}, {Bennett},
  {Bhattacharya}, {Donachie}, {Freeman}, {Fukui}, {Hirao}, {Inayama}, {Itow},
  {Koshimoto}, {Li}, {Ling}, {Masuda}, {Matsubara}, {Muraki}, {Nagakane},
  {Nishioka}, {Ohnishi}, {Oyokawa}, {Rattenbury}, {Saito}, {Sharan},
  {Sullivan}, {Sumi}, {Suzuki}, {Tristram}, {Wakiyama}, {Yonehara}, {MOA
  Group}, {Choi}, {Park}, {Jung}, {Shin}, {Albrow}, {Park}, {Kim}, {Lee},
  {Cha}, {Kim}, {Lee}, {KMTNet Group}, {Dominik}, {J{\o}rgensen}, {Andersen},
  {Bramich}, {Burgdorf}, {Ciceri}, {D'Ago}, {Evans}, {Figuera Jaimes}, {Gu},
  {Hinse}, {Kains}, {Kerins}, {Korhonen}, {Kuffmeier}, {Mancini}, {Popovas},
  {Rabus}, {Rahvar}, {Rasmussen}, {Scarpetta}, {Skottfelt}, {Snodgrass},
  {Southworth}, {Surdej}, {Unda-Sanzana}, {von Essen}, {Wang}, {Wertz},
  {MiNDSTEp}, {Maoz}, {Friedmann}, {Kaspi}, \& {Wise Group}}]{Bozza16}
{Bozza}, V., {Shvartzvald}, Y., {Udalski}, A., {et~al.} 2016, \apj, 820, 79,
  \dodoi{10.3847/0004-637X/820/1/79}

\bibitem[{{Calchi Novati} {et~al.}(2015{\natexlab{a}}){Calchi Novati}, {Gould},
  {Udalski}, {Menzies}, {Bond}, {Shvartzvald}, {Street}, {Hundertmark},
  {Beichman}, {Yee}, {Carey}, {Poleski}, {Skowron}, {Koz{\l}owski}, {Mr{\'o}z},
  {Pietrukowicz}, {Pietrzy{\'n}ski}, {Szyma{\'n}ski}, {Soszy{\'n}ski},
  {Ulaczyk}, {Wyrzykowski}, {OGLE Collaboration}, {Albrow}, {Beaulieu},
  {Caldwell}, {Cassan}, {Coutures}, {Danielski}, {Dominis Prester},
  {Donatowicz}, {Lon{\v c}ari{\'c}}, {McDougall}, {Morales}, {Ranc}, {Zhu},
  {PLANET Collaboration}, {Abe}, {Barry}, {Bennett}, {Bhattacharya},
  {Fukunaga}, {Inayama}, {Koshimoto}, {Namba}, {Sumi}, {Suzuki}, {Tristram},
  {Wakiyama}, {Yonehara}, {MOA Collaboration}, {Maoz}, {Kaspi}, {Friedmann},
  {Wise Group}, {Bachelet}, {Figuera Jaimes}, {Bramich}, {Tsapras}, {Horne},
  {Snodgrass}, {Wambsganss}, {Steele}, {Kains}, {RoboNet Collaboration},
  {Bozza}, {Dominik}, {J{\o}rgensen}, {Alsubai}, {Ciceri}, {D'Ago},
  {Haugb{\o}lle}, {Hessman}, {Hinse}, {Juncher}, {Korhonen}, {Mancini},
  {Popovas}, {Rabus}, {Rahvar}, {Scarpetta}, {Schmidt}, {Skottfelt},
  {Southworth}, {Starkey}, {Surdej}, {Wertz}, {Zarucki}, {MiNDSTEp Consortium},
  {Gaudi}, {Pogge}, {DePoy}, \& {{$\mu$}FUN Collaboration}}]{CalchiNovati15a}
{Calchi Novati}, S., {Gould}, A., {Udalski}, A., {et~al.} 2015{\natexlab{a}},
  \apj, 804, 20, \dodoi{10.1088/0004-637X/804/1/20}

\bibitem[{{Calchi Novati} {et~al.}(2015{\natexlab{b}}){Calchi Novati}, {Gould},
  {Yee}, {Beichman}, {Bryden}, {Carey}, {Fausnaugh}, {Gaudi}, {Henderson},
  {Pogge}, {Shvartzvald}, {Wibking}, {Zhu}, {Spitzer Team}, {Udalski},
  {Poleski}, {Pawlak}, {Szyma{\'n}ski}, {Skowron}, {Mr{\'o}z}, {Koz{\l}owski},
  {Wyrzykowski}, {Pietrukowicz}, {Pietrzy{\'n}ski}, {Soszy{\'n}ski}, {Ulaczyk},
  \& {OGLE Group}}]{CalchiNovati15b}
{Calchi Novati}, S., {Gould}, A., {Yee}, J.~C., {et~al.} 2015{\natexlab{b}},
  \apj, 814, 92, \dodoi{10.1088/0004-637X/814/2/92}

\bibitem[{{Casagrande} {et~al.}(2010){Casagrande}, {Ram{\'{\i}}rez},
  {Mel{\'e}ndez}, {Bessell}, \& {Asplund}}]{Casagrande10}
{Casagrande}, L., {Ram{\'{\i}}rez}, I., {Mel{\'e}ndez}, J., {Bessell}, M., \&
  {Asplund}, M. 2010, \aap, 512, A54, \dodoi{10.1051/0004-6361/200913204}

\bibitem[{{Chabrier}(2003)}]{Chabrier03}
{Chabrier}, G. 2003, \pasp, 115, 763, \dodoi{10.1086/376392}

\bibitem[{{Chung} {et~al.}(2017){Chung}, {Zhu}, {Udalski}, {Lee}, {Ryu},
  {Jung}, {Shin}, {Yee}, {Hwang}, {Gould}, {and}, {Albrow}, {Cha}, {Han},
  {Kim}, {Kim}, {Kim}, {Kim}, {Lee}, {Park}, {Pogge}, {KMTNet Collaboration},
  {Poleski}, {Mr{\'o}z}, {Pietrukowicz}, {Skowron}, {Szyma{\'n}ski},
  {Soszy{\'n}ski}, {Koz{\l}owski}, {Ulaczyk}, {Pawlak}, {OGLE Collaboration},
  {Beichman}, {Bryden}, {Calchi Novati}, {Carey}, {Fausnaugh}, {Gaudi},
  {Henderson}, {Shvartzvald}, {Wibking}, \& {The Spitzer Team}}]{Chung17}
{Chung}, S.-J., {Zhu}, W., {Udalski}, A., {et~al.} 2017, \apj, 838, 154,
  \dodoi{10.3847/1538-4357/aa67fa}

\bibitem[{{Claret} \& {Bloemen}(2011)}]{Claret11}
{Claret}, A., \& {Bloemen}, S. 2011, \aap, 529, A75,
  \dodoi{10.1051/0004-6361/201116451}

\bibitem[{{Dong} {et~al.}(2006){Dong}, {DePoy}, {Gaudi}, {Gould}, {Han},
  {Park}, {Pogge}, {MuFun Collaboration}, {Udalski}, {Szewczyk}, {Kubiak},
  {Szyma{\'n}ski}, {Pietrzy{\'n}ski}, {Soszy{\'n}ski}, {Wyrzykowski},
  {{\.Z}ebru{\'n}}, \& {OGLE Collaboration}}]{Dong06}
{Dong}, S., {DePoy}, D.~L., {Gaudi}, B.~S., {et~al.} 2006, \apj, 642, 842,
  \dodoi{10.1086/501224}

\bibitem[{{Gould}(1994)}]{Gould94}
{Gould}, A. 1994, \apjl, 421, L75, \dodoi{10.1086/187191}

\bibitem[{{Gould}(2000)}]{Gould00}
---. 2000, \apj, 542, 785, \dodoi{10.1086/317037}

\bibitem[{{Gould}(2008)}]{Gould08}
---. 2008, \apj, 681, 1593, \dodoi{10.1086/588601}

\bibitem[{{Gould} {et~al.}(2006){Gould}, {Udalski}, {An}, {Bennett}, {Zhou},
  {Dong}, {Rattenbury}, {Gaudi}, {Yock}, {Bond}, {Christie}, {Horne},
  {Anderson}, {Stanek}, {DePoy}, {Han}, {McCormick}, {Park}, {Pogge},
  {Poindexter}, {Soszy{\'n}ski}, {Szyma{\'n}ski}, {Kubiak}, {Pietrzy{\'n}ski},
  {Szewczyk}, {Wyrzykowski}, {Ulaczyk}, {Paczy{\'n}ski}, {Bramich},
  {Snodgrass}, {Steele}, {Burgdorf}, {Bode}, {Botzler}, {Mao}, \&
  {Swaving}}]{Gould06}
{Gould}, A., {Udalski}, A., {An}, D., {et~al.} 2006, \apjl, 644, L37,
  \dodoi{10.1086/505421}

\bibitem[{{Han} \& {Gould}(1995)}]{Han95}
{Han}, C., \& {Gould}, A. 1995, \apj, 447, 53, \dodoi{10.1086/175856}

\bibitem[{{Han} \& {Gould}(2003)}]{Han03}
---. 2003, \apj, 592, 172, \dodoi{10.1086/375706}

\bibitem[{{Han} {et~al.}(2016){Han}, {Udalski}, {Gould}, {Zhu}, {Street},
  {Yee}, {Beichman}, {Bryden}, {Calchi Novati}, {Carey}, {Fausnaugh}, {Gaudi},
  {Henderson}, {Shvartzvald}, {Wibking}, {Spitzer Microlensing Team},
  {Szyma{\'n}ski}, {Soszy{\'n}ski}, {Skowron}, {Mr{\'o}z}, {Poleski},
  {Pietrukowicz}, {Koz{\l}owski}, {Ulaczyk}, {Wyrzykowski}, {Pawlak}, {OGLE
  Collaboration}, {Tsapras}, {Hundertmark}, {Bachelet}, {Dominik}, {Bramich},
  {Cassan}, {Figuera Jaimes}, {Horne}, {Ranc}, {Schmidt}, {Snodgrass},
  {Wambsganss}, {Steele}, {Menzies}, {Mao}, {RoboNet Collaboration}, {Bozza},
  {J{\o}rgensen}, {Alsubai}, {Ciceri}, {D'Ago}, {Haugb{\o}lle}, {Hessman},
  {Hinse}, {Juncher}, {Korhonen}, {Mancini}, {Popovas}, {Rabus}, {Rahvar},
  {Scarpetta}, {Skottfelt}, {Southworth}, {Starkey}, {Surdej}, {Wertz},
  {Zarucki}, {MiNDSTEp Consortium}, {Pogge}, {DePoy}, \& {{$\mu$}FUN
  Collaboration}}]{Han16}
{Han}, C., {Udalski}, A., {Gould}, A., {et~al.} 2016, \apj, 828, 53,
  \dodoi{10.3847/0004-637X/828/1/53}

\bibitem[{{Han} {et~al.}(2017){Han}, {Udalski}, {Gould}, {Zhu}, {and},
  {Szyma{\'n}ski}, {Soszy{\'n}ski}, {Skowron}, {Mr{\'o}z}, {Poleski},
  {Pietrukowicz}, {Koz{\l}owski}, {Ulaczyk}, {Pawlak}, {OGLE Collaboration},
  {Yee}, {Beichman}, {Calchi Novati}, {Carey}, {Bryden}, {Fausnaugh}, {Gaudi},
  {Henderson}, {Shvartzvald}, {Wibking}, \& {The Spitzer Microlensing
  Team}}]{Han17}
---. 2017, \apj, 834, 82, \dodoi{10.3847/1538-4357/834/1/82}

\bibitem[{{Han} {et~al.}(2018){Han}, {Calchi Novati}, {Udalski}, {Lee},
  {Gould}, {Bozza}, {Mr{\'o}z}, {Pietrukowicz}, {Skowron}, {Szyma{\'n}ski},
  {Poleski}, {Soszy{\'n}ski}, {Koz{\l}owski}, {Ulaczyk}, {Pawlak}, {Rybicki},
  {Iwanek}, {Albrow}, {Chung}, {Hwang}, {Jung}, {Ryu}, {Shin}, {Shvartzvald},
  {Yee}, {Zang}, {Zhu}, {Cha}, {Kim}, {Kim}, {Kim}, {Lee}, {Lee}, {Park},
  {Pogge}, {Kim}, {Beichman}, {Bryden}, {Carey}, {Gaudi}, {Henderson},
  {Dominik}, {Helling}, {Hundertmark}, {J{\o}rgensen}, {Longa-Pe{\~n}a},
  {Lowry}, {Sajadian}, {Burgdorf}, {Campbell-White}, {Ciceri}, {Evans},
  {Haikala}, {Hinse}, {Rahvar}, {Rabus}, \& {Snodgrass}}]{Han18}
{Han}, C., {Calchi Novati}, S., {Udalski}, A., {et~al.} 2018, ArXiv e-prints.
\newblock \doarXiv{1802.10196}

\bibitem[{{Jung} {et~al.}(2018){Jung}, {Udalski}, {Gould}, {Ryu}, {Yee}, {Han},
  {Albrow}, {Lee}, {Kim}, {Hwang}, {Chung}, {Shin}, {Zhu}, {Cha}, {Kim}, {Lee},
  {Park}, {Lee}, {Kim}, {Pogge}, {Szyma{\'n}ski}, {Mr{\'o}z}, {Poleski},
  {Skowron}, {Pietrukowicz}, {Soszy{\'n}ski}, {Koz{\l}owski}, {Ulaczyk},
  {Pawlak}, \& {Rybicki}}]{Jung18}
{Jung}, Y.~K., {Udalski}, A., {Gould}, A., {et~al.} 2018, ArXiv e-prints.
\newblock \doarXiv{1803.05095}

\bibitem[{{Kayser} {et~al.}(1986){Kayser}, {Refsdal}, \& {Stabell}}]{Kayser86}
{Kayser}, R., {Refsdal}, S., \& {Stabell}, R. 1986, \aap, 166, 36

\bibitem[{{Mr{\'o}z} {et~al.}(2017){Mr{\'o}z}, {Udalski}, {Skowron}, {Poleski},
  {Koz{\l}owski}, {Szyma{\'n}ski}, {Soszy{\'n}ski}, {Wyrzykowski},
  {Pietrukowicz}, {Ulaczyk}, {Skowron}, \& {Pawlak}}]{Mroz17}
{Mr{\'o}z}, P., {Udalski}, A., {Skowron}, J., {et~al.} 2017, \nat, 548, 183,
  \dodoi{10.1038/nature23276}

\bibitem[{{Mr{\'o}z} {et~al.}(2018){Mr{\'o}z}, {Ryu}, {Skowron}, {Udalski},
  {Gould}, {Szyma{\'n}ski}, {Soszy{\'n}ski}, {Poleski}, {Pietrukowicz},
  {Koz{\l}owski}, {Pawlak}, {Ulaczyk}, {OGLE Collaboration}, {Albrow}, {Chung},
  {Jung}, {Han}, {Hwang}, {Shin}, {Yee}, {Zhu}, {Cha}, {Kim}, {Kim}, {Kim},
  {Lee}, {Lee}, {Lee}, {Park}, {Pogge}, \& {KMTNet Collaboration}}]{Mroz18}
{Mr{\'o}z}, P., {Ryu}, Y.-H., {Skowron}, J., {et~al.} 2018, \aj, 155, 121,
  \dodoi{10.3847/1538-3881/aaaae9}

\bibitem[{{Nataf} {et~al.}(2013){Nataf}, {Gould}, {Fouqu{\'e}}, {Gonzalez},
  {Johnson}, {Skowron}, {Udalski}, {Szyma{\'n}ski}, {Kubiak},
  {Pietrzy{\'n}ski}, {Soszy{\'n}ski}, {Ulaczyk}, {Wyrzykowski}, \&
  {Poleski}}]{Nataf13}
{Nataf}, D.~M., {Gould}, A., {Fouqu{\'e}}, P., {et~al.} 2013, \apj, 769, 88,
  \dodoi{10.1088/0004-637X/769/2/88}

\bibitem[{{Pejcha} \& {Heyrovsk{\'y}}(2009)}]{Pejcha09}
{Pejcha}, O., \& {Heyrovsk{\'y}}, D. 2009, \apj, 690, 1772,
  \dodoi{10.1088/0004-637X/690/2/1772}

\bibitem[{{Penny} {et~al.}(2016){Penny}, {Henderson}, \& {Clanton}}]{Penny16}
{Penny}, M.~T., {Henderson}, C.~B., \& {Clanton}, C. 2016, \apj, 830, 150,
  \dodoi{10.3847/0004-637X/830/2/150}

\bibitem[{{Refsdal}(1966)}]{Refsdal66}
{Refsdal}, S. 1966, \mnras, 134, 315, \dodoi{10.1093/mnras/134.3.315}

\bibitem[{{Ryu} {et~al.}(2018){Ryu}, {Yee}, {Udalski}, {Bond}, {Shvartzvald},
  {Zang}, {Figuera Jaimes}, {J{\o}rgensen}, {Zhu}, {Huang}, {Jung}, {Albrow},
  {Chung}, {Gould}, {Han}, {Hwang}, {Shin}, {Cha}, {Kim}, {Kim}, {Kim}, {Lee},
  {Lee}, {Lee}, {Park}, {Pogge}, {KMTNet Collaboration}, {Calchi Novati},
  {Carey}, {Henderson}, {Beichman}, {Gaudi}, {Spitzer team}, {Mr{\'o}z},
  {Poleski}, {Skowron}, {Szyma{\'n}ski}, {Soszy{\'n}ski}, {Koz{\l}owski},
  {Pietrukowicz}, {Ulaczyk}, {Pawlak}, {OGLE Collaboration}, {Abe}, {Asakura},
  {Barry}, {Bennett}, {Bhattacharya}, {Donachie}, {Evans}, {Fukui}, {Hirao},
  {Itow}, {Kawasaki}, {Koshimoto}, {Li}, {Ling}, {Masuda}, {Matsubara},
  {Miyazaki}, {Muraki}, {Nagakane}, {Ohnishi}, {Ranc}, {Rattenbury}, {Saito},
  {Sharan}, {Sullivan}, {Sumi}, {Suzuki}, {Tristram}, {Yamada}, {Yamada},
  {Yonehara}, {MOA Collaboration}, {Bryden}, {Howell}, {Jacklin}, {UKIRT
  Microlensing Team}, {Penny}, {Mao}, {Fouqu{\'e}}, {Wang}, {CFHT-K2C9
  Microlensing Survey group}, {Street}, {Tsapras}, {Hundertmark}, {Bachelet},
  {Dominik}, {Li}, {Cross}, {Cassan}, {Horne}, {Schmidt}, {Wambsganss}, {Ment},
  {Maoz}, {Snodgrass}, {Steele}, {RoboNet Team}, {Bozza}, {Burgdorf}, {Ciceri},
  {D{\rsquo}Ago}, {Evans}, {Hinse}, {Kerins}, {Kokotanekova}, {Longa},
  {MacKenzie}, {Popovas}, {Rabus}, {Rahvar}, {Sajadian}, {Skottfelt},
  {Southworth}, {von Essen}, \& {MiNDSTEp Team}}]{Ryu18}
{Ryu}, Y.-H., {Yee}, J.~C., {Udalski}, A., {et~al.} 2018, \aj, 155, 40,
  \dodoi{10.3847/1538-3881/aa9be4}

\bibitem[{{Schneider} \& {Weiss}(1986)}]{Schneider86}
{Schneider}, P., \& {Weiss}, A. 1986, \aap, 164, 237

\bibitem[{{Shin} {et~al.}(2017){Shin}, {Udalski}, {Yee}, {Calchi Novati},
  {Han}, {Skowron}, {Mr{\'o}z}, {Soszy{\'n}ski}, {Poleski}, {Szyma{\'n}ski},
  {Koz{\l}owski}, {Pietrukowicz}, {Ulaczyk}, {Pawlak}, {OGLE Collaboration},
  {Albrow}, {Gould}, {Chung}, {Hwang}, {Jung}, {Ryu}, {Zhu}, {Cha}, {Kim},
  {Kim}, {Kim}, {Lee}, {Lee}, {Park}, {Pogge}, {KMTNet Group}, {Beichman},
  {Bryden}, {Carey}, {Gaudi}, {Henderson}, {Shvartzvald}, \& {Spitzer
  Team}}]{Shin17}
{Shin}, I.-G., {Udalski}, A., {Yee}, J.~C., {et~al.} 2017, \aj, 154, 176,
  \dodoi{10.3847/1538-3881/aa8a74}

\bibitem[{{Shvartzvald} {et~al.}(2015){Shvartzvald}, {Udalski}, {Gould}, {Han},
  {Bozza}, {Friedmann}, {Hundertmark}, {and}, {Beichman}, {Bryden}, {Calchi
  Novati}, {Carey}, {Fausnaugh}, {Gaudi}, {Henderson}, {Kerr}, {Pogge},
  {Varricatt}, {Wibking}, {Yee}, {Zhu}, {Spitzer Team}, {Poleski}, {Pawlak},
  {Szyma{\'n}ski}, {Skowron}, {Mr{\'o}z}, {Koz{\l}owski}, {Wyrzykowski},
  {Pietrukowicz}, {Pietrzy{\'n}ski}, {Soszy{\'n}ski}, {Ulaczyk}, {OGLE Group},
  {Choi}, {Park}, {Jung}, {Shin}, {Albrow}, {Park}, {Kim}, {Lee}, {Cha}, {Kim},
  {Lee}, {KMTNet Group}, {Maoz}, {Kaspi}, {Wise Group}, {Street}, {Tsapras},
  {Bachelet}, {Dominik}, {Bramich}, {Horne}, {Snodgrass}, {Steele}, {Menzies},
  {Figuera Jaimes}, {Wambsganss}, {Schmidt}, {Cassan}, {Ranc}, {Mao}, {Dong},
  {RoboNet}, {D'Ago}, {Scarpetta}, {Verma}, {J{\o}rgensen}, {Kerins},
  {Skottfelt}, \& {MiNDSTEp}}]{Shvartzvald15}
{Shvartzvald}, Y., {Udalski}, A., {Gould}, A., {et~al.} 2015, \apj, 814, 111,
  \dodoi{10.1088/0004-637X/814/2/111}

\bibitem[{{Shvartzvald} {et~al.}(2016{\natexlab{a}}){Shvartzvald}, {Maoz},
  {Udalski}, {Sumi}, {Friedmann}, {Kaspi}, {Poleski}, {Szyma{\'n}ski},
  {Skowron}, {Koz{\l}owski}, {Wyrzykowski}, {Mr{\'o}z}, {Pietrukowicz},
  {Pietrzy{\'n}ski}, {Soszy{\'n}ski}, {Ulaczyk}, {Abe}, {Barry}, {Bennett},
  {Bhattacharya}, {Bond}, {Freeman}, {Inayama}, {Itow}, {Koshimoto}, {Ling},
  {Masuda}, {Fukui}, {Matsubara}, {Muraki}, {Ohnishi}, {Rattenbury}, {Saito},
  {Sullivan}, {Suzuki}, {Tristram}, {Wakiyama}, \& {Yonehara}}]{Shvartzvald16b}
{Shvartzvald}, Y., {Maoz}, D., {Udalski}, A., {et~al.} 2016{\natexlab{a}},
  \mnras, 457, 4089, \dodoi{10.1093/mnras/stw191}

\bibitem[{{Shvartzvald} {et~al.}(2016{\natexlab{b}}){Shvartzvald}, {Li},
  {Udalski}, {Gould}, {Sumi}, {Street}, {Calchi Novati}, {Hundertmark},
  {Bozza}, {Beichman}, {Bryden}, {Carey}, {Drummond}, {Fausnaugh}, {Gaudi},
  {Henderson}, {Tan}, {Wibking}, {Pogge}, {Yee}, {Zhu}, {(Spitzer Team},
  {Tsapras}, {Bachelet}, {Dominik}, {Bramich}, {Cassan}, {Figuera Jaimes},
  {Horne}, {Ranc}, {Schmidt}, {Snodgrass}, {Wambsganss}, {Steele}, {Menzies},
  {Mao}, {(RoboNet}, {Poleski}, {Pawlak}, {Szyma{\'n}ski}, {Skowron},
  {Mr{\'o}z}, {Koz{\l}owski}, {Wyrzykowski}, {Pietrukowicz}, {Soszy{\'n}ski},
  {Ulaczyk}, {(OGLE Group}, {Abe}, {Asakura}, {Barry}, {Bennett},
  {Bhattacharya}, {Bond}, {Freeman}, {Hirao}, {Itow}, {Koshimoto}, {Li},
  {Ling}, {Masuda}, {Fukui}, {Matsubara}, {Muraki}, {Nagakane}, {Nishioka},
  {Ohnishi}, {Oyokawa}, {Rattenbury}, {Saito}, {Sharan}, {Sullivan}, {Suzuki},
  {Tristram}, {Yonehara}, {(MOA Group}, {J{\o}rgensen}, {Burgdorf}, {Ciceri},
  {D'Ago}, {Evans}, {Hinse}, {Kains}, {Kerins}, {Korhonen}, {Mancini},
  {Popovas}, {Rabus}, {Rahvar}, {Scarpetta}, {Skottfelt}, {Southworth},
  {Peixinho}, {Verma}, {(MiNDSTEp}, {Sbarufatti}, {Kennea}, {Gehrels}, \&
  {(Swift}}]{Shvartzvald16}
{Shvartzvald}, Y., {Li}, Z., {Udalski}, A., {et~al.} 2016{\natexlab{b}}, \apj,
  831, 183, \dodoi{10.3847/0004-637X/831/2/183}

\bibitem[{{Shvartzvald} {et~al.}(2017){Shvartzvald}, {Yee}, {Calchi Novati},
  {Gould}, {Lee}, {Beichman}, {Bryden}, {Carey}, {Gaudi}, {Henderson}, {Zhu},
  {Spitzer Team}, {Albrow}, {Cha}, {Chung}, {Han}, {Hwang}, {Jung}, {Kim},
  {Kim}, {Kim}, {Lee}, {Park}, {Pogge}, {Ryu}, {Shin}, \& {KMTNet
  Group}}]{Shvartzvald17}
{Shvartzvald}, Y., {Yee}, J.~C., {Calchi Novati}, S., {et~al.} 2017, \apjl,
  840, L3, \dodoi{10.3847/2041-8213/aa6d09}

\bibitem[{{Skowron} {et~al.}(2016){Skowron}, {Udalski}, {Koz{\l}owski},
  {Szyma{\'n}ski}, {Mr{\'o}z}, {Wyrzykowski}, {Poleski}, {Pietrukowicz},
  {Ulaczyk}, {Pawlak}, \& {Soszy{\'n}ski}}]{Skowron16}
{Skowron}, J., {Udalski}, A., {Koz{\l}owski}, S., {et~al.} 2016, \actaa, 66, 1.
\newblock \doarXiv{1604.01966}

\bibitem[{{Spergel} {et~al.}(2013){Spergel}, {Gehrels}, {Breckinridge},
  {Donahue}, {Dressler}, {Gaudi}, {Greene}, {Guyon}, {Hirata}, {Kalirai},
  {Kasdin}, {Moos}, {Perlmutter}, {Postman}, {Rauscher}, {Rhodes}, {Wang},
  {Weinberg}, {Centrella}, {Traub}, {Baltay}, {Colbert}, {Bennett},
  {Kiessling}, {Macintosh}, {Merten}, {Mortonson}, {Penny}, {Rozo},
  {Savransky}, {Stapelfeldt}, {Zu}, {Baker}, {Cheng}, {Content}, {Dooley},
  {Foote}, {Goullioud}, {Grady}, {Jackson}, {Kruk}, {Levine}, {Melton},
  {Peddie}, {Ruffa}, \& {Shaklan}}]{Spergel13}
{Spergel}, D., {Gehrels}, N., {Breckinridge}, J., {et~al.} 2013, ArXiv
  e-prints.
\newblock \doarXiv{1305.5422}

\bibitem[{{Street} {et~al.}(2016){Street}, {Udalski}, {Calchi Novati},
  {Hundertmark}, {Zhu}, {Gould}, {Yee}, {Tsapras}, {Bennett}, {RoboNet
  Project}, {Consortium}, {J{\o}rgensen}, {Dominik}, {Andersen}, {Bachelet},
  {Bozza}, {Bramich}, {Burgdorf}, {Cassan}, {Ciceri}, {D'Ago}, {Dong}, {Evans},
  {Gu}, {Harkonnen}, {Hinse}, {Horne}, {Figuera Jaimes}, {Kains}, {Kerins},
  {Korhonen}, {Kuffmeier}, {Mancini}, {Menzies}, {Mao}, {Peixinho}, {Popovas},
  {Rabus}, {Rahvar}, {Ranc}, {Tronsgaard Rasmussen}, {Scarpetta}, {Schmidt},
  {Skottfelt}, {Snodgrass}, {Southworth}, {Steele}, {Surdej}, {Unda-Sanzana},
  {Verma}, {von Essen}, {Wambsganss}, {Wang}, {Wertz}, {OGLE Project},
  {Poleski}, {Pawlak}, {Szyma{\'n}ski}, {Skowron}, {Mr{\'o}z}, {Koz{\l}owski},
  {Wyrzykowski}, {Pietrukowicz}, {Pietrzy{\'n}ski}, {Soszy{\'n}ski}, {Ulaczyk},
  {Spitzer Team}, {Beichman}, {Bryden}, {Carey}, {Gaudi}, {Henderson}, {Pogge},
  {Shvartzvald}, {MOA Collaboration}, {Abe}, {Asakura}, {Bhattacharya}, {Bond},
  {Donachie}, {Freeman}, {Fukui}, {Hirao}, {Inayama}, {Itow}, {Koshimoto},
  {Li}, {Ling}, {Masuda}, {Matsubara}, {Muraki}, {Nagakane}, {Nishioka},
  {Ohnishi}, {Oyokawa}, {Rattenbury}, {Saito}, {Sharan}, {Sullivan}, {Sumi},
  {Suzuki}, {Tristram}, {Wakiyama}, {Yonehara}, {KMTNet Modeling Team}, {Han},
  {Choi}, {Park}, {Jung}, \& {Shin}}]{Street16}
{Street}, R.~A., {Udalski}, A., {Calchi Novati}, S., {et~al.} 2016, \apj, 819,
  93, \dodoi{10.3847/0004-637X/819/2/93}

\bibitem[{{Sumi} {et~al.}(2011){Sumi}, {Kamiya}, {Bennett}, {Bond}, {Abe},
  {Botzler}, {Fukui}, {Furusawa}, {Hearnshaw}, {Itow}, {Kilmartin}, {Korpela},
  {Lin}, {Ling}, {Masuda}, {Matsubara}, {Miyake}, {Motomura}, {Muraki},
  {Nagaya}, {Nakamura}, {Ohnishi}, {Okumura}, {Perrott}, {Rattenbury}, {Saito},
  {Sako}, {Sullivan}, {Sweatman}, {Tristram}, {Udalski}, {Szyma{\'n}ski},
  {Kubiak}, {Pietrzy{\'n}ski}, {Poleski}, {Soszy{\'n}ski}, {Wyrzykowski},
  {Ulaczyk}, \& {Microlensing Observations in Astrophysics (MOA)
  Collaboration}}]{Sumi11}
{Sumi}, T., {Kamiya}, K., {Bennett}, D.~P., {et~al.} 2011, \nat, 473, 349,
  \dodoi{10.1038/nature10092}

\bibitem[{{Suzuki} {et~al.}(2016){Suzuki}, {Bennett}, {Sumi}, {Bond}, {Rogers},
  {Abe}, {Asakura}, {Bhattacharya}, {Donachie}, {Freeman}, {Fukui}, {Hirao},
  {Itow}, {Koshimoto}, {Li}, {Ling}, {Masuda}, {Matsubara}, {Muraki},
  {Nagakane}, {Onishi}, {Oyokawa}, {Rattenbury}, {Saito}, {Sharan}, {Shibai},
  {Sullivan}, {Tristram}, {Yonehara}, \& {MOA Collaboration}}]{Suzuki16}
{Suzuki}, D., {Bennett}, D.~P., {Sumi}, T., {et~al.} 2016, \apj, 833, 145,
  \dodoi{10.3847/1538-4357/833/2/145}

\bibitem[{{Udalski}(2003)}]{Udalski03}
{Udalski}, A. 2003, \actaa, 53, 291

\bibitem[{{Udalski} {et~al.}(2015{\natexlab{a}}){Udalski}, {Szyma{\'n}ski}, \&
  {Szyma{\'n}ski}}]{Udalski15a}
{Udalski}, A., {Szyma{\'n}ski}, M.~K., \& {Szyma{\'n}ski}, G.
  2015{\natexlab{a}}, \actaa, 65, 1.
\newblock \doarXiv{1504.05966}

\bibitem[{{Udalski} {et~al.}(2015{\natexlab{b}}){Udalski}, {Yee}, {Gould},
  {Carey}, {Zhu}, {Skowron}, {Koz{\l}owski}, {Poleski}, {Pietrukowicz},
  {Pietrzy{\'n}ski}, {Szyma{\'n}ski}, {Mr{\'o}z}, {Soszy{\'n}ski}, {Ulaczyk},
  {Wyrzykowski}, {Han}, {Calchi Novati}, \& {Pogge}}]{Udalski15b}
{Udalski}, A., {Yee}, J.~C., {Gould}, A., {et~al.} 2015{\natexlab{b}}, \apj,
  799, 237, \dodoi{10.1088/0004-637X/799/2/237}

\bibitem[{{Udalski} {et~al.}(2018){Udalski}, {Han}, {Bozza}, {Gould}, {Bond},
  {and}, {Mr{\'o}z}, {Skowron}, {Wyrzykowski}, {Szyma{\'n}ski},
  {Soszy{\'n}ski}, {Ulaczyk}, {Poleski}, {Pietrukowicz}, {Koz{\l}owski}, {The
  OGLE Collaboration}, {Abe}, {Barry}, {Bennett}, {Bhattacharya}, {Donachie},
  {Evans}, {Fukui}, {Hirao}, {Itow}, {Kawasaki}, {Koshimoto}, {Li}, {Ling},
  {Masuda}, {Matsubara}, {Miyazaki}, {Munakata}, {Muraki}, {Nagakane},
  {Ohnishi}, {Ranc}, {Rattenbury}, {Saito}, {Sharan}, {Sullivan}, {Sumi},
  {Suzuki}, {Tristram}, {Yamada}, {Yonehara}, {The MOA Collaboration},
  {Street}, {Tsapras}, {Bachelet}, {Bramich}, {D{\'A}go}, {Dominik}, {Figuera
  Jaimes}, {Horne}, {Hundertmark}, {Kains}, {Menzies}, {Schmidt}, {Snodgrass},
  {Steele}, {Wambsganss}, {Robonet Collaboration}, {Pogge}, {Jung}, {Shin},
  {Yee}, {Kim}, {The {$\mu$}Fun Collaboration}, {Beichman}, {Carey}, {Calchi
  Novati}, {Zhu}, \& {The Spitzer Team}}]{Udalski18}
{Udalski}, A., {Han}, C., {Bozza}, V., {et~al.} 2018, \apj, 853, 70,
  \dodoi{10.3847/1538-4357/aaa295}

\bibitem[{{Wambsganss}(1997)}]{Wambsganss97}
{Wambsganss}, J. 1997, \mnras, 284, 172, \dodoi{10.1093/mnras/284.1.172}

\bibitem[{{Wang} {et~al.}(2017){Wang}, {Zhu}, {Mao}, {Bond}, {Gould},
  {Udalski}, {Sumi}, {Bozza}, {Ranc}, {Cassan}, {Yee}, {Han}, {Abe}, {Asakura},
  {Barry}, {Bennett}, {Bhattacharya}, {Donachie}, {Evans}, {Fukui}, {Hirao},
  {Itow}, {Kawasaki}, {Koshimoto}, {Li}, {Ling}, {Masuda}, {Matsubara},
  {Miyazaki}, {Muraki}, {Nagakane}, {Ohnishi}, {Rattenbury}, {Saito}, {Sharan},
  {Shibai}, {Sullivan}, {Suzuki}, {Tristram}, {Yamada}, {Yonehara}, {MOA
  Collaboration}, {Koz{\L}owski}, {Mr{\'o}z}, {Pawlak}, {Pietrukowicz},
  {Poleski}, {Skowron}, {Soszy{\'n}ski}, {Szyma{\'n}ski}, {Ulaczyk}, {OGLE
  Collaboration}, {Beichman}, {Bryden}, {Calchi Novati}, {Carey}, {Fausnaugh},
  {Gaudi}, {Henderson}, {Shvartzvald}, {Wibking}, {Spitzer Team}, {Albrow},
  {Chung}, {Hwang}, {Jung}, {Ryu}, {Shin}, {Cha}, {Kim}, {Kim}, {Kim}, {Lee},
  {Lee}, {Park}, {Pogge}, {KMTNet Collaboration}, {Street}, {Tsapras},
  {Hundertmark}, {Bachelet}, {Dominik}, {Horne}, {Figuera Jaimes},
  {Wambsganss}, {Bramich}, {Schmidt}, {Snodgrass}, {Steele}, {Menzies}, \&
  {RoboNet Collaboration}}]{Wang17}
{Wang}, T., {Zhu}, W., {Mao}, S., {et~al.} 2017, \apj, 845, 129,
  \dodoi{10.3847/1538-4357/aa813b}

\bibitem[{{Wozniak}(2000)}]{Wozniak00}
{Wozniak}, P.~R. 2000, \actaa, 50, 421

\bibitem[{{Wyrzykowski} {et~al.}(2016){Wyrzykowski}, {Kostrzewa-Rutkowska},
  {Skowron}, {Rybicki}, {Mr{\'o}z}, {Koz{\l}owski}, {Udalski}, {Szyma{\'n}ski},
  {Pietrzy{\'n}ski}, {Soszy{\'n}ski}, {Ulaczyk}, {Pietrukowicz}, {Poleski},
  {Pawlak}, {I{\l}kiewicz}, \& {Rattenbury}}]{Wyrzykowski16}
{Wyrzykowski}, {\L}., {Kostrzewa-Rutkowska}, Z., {Skowron}, J., {et~al.} 2016,
  \mnras, 458, 3012, \dodoi{10.1093/mnras/stw426}

\bibitem[{{Yee} {et~al.}(2012){Yee}, {Shvartzvald}, {Gal-Yam}, {Bond},
  {Udalski}, {Koz{\l}owski}, {Han}, {Gould}, {Skowron}, {Suzuki}, {Abe},
  {Bennett}, {Botzler}, {Chote}, {Freeman}, {Fukui}, {Furusawa}, {Itow},
  {Kobara}, {Ling}, {Masuda}, {Matsubara}, {Miyake}, {Muraki}, {Ohmori},
  {Ohnishi}, {Rattenbury}, {Saito}, {Sullivan}, {Sumi}, {Suzuki}, {Sweatman},
  {Takino}, {Tristram}, {Wada}, {MOA Collaboration}, {Szyma{\'n}ski}, {Kubiak},
  {Pietrzy{\'n}ski}, {Soszy{\'n}ski}, {Poleski}, {Ulaczyk}, {Wyrzykowski},
  {Pietrukowicz}, {OGLE Collaboration}, {Allen}, {Almeida}, {Batista}, {Bos},
  {Christie}, {DePoy}, {Dong}, {Drummond}, {Finkelman}, {Gaudi}, {Gorbikov},
  {Henderson}, {Higgins}, {Jablonski}, {Kaspi}, {Manulis}, {Maoz}, {McCormick},
  {McGregor}, {Monard}, {Moorhouse}, {Mu{\~n}oz}, {Natusch}, {Ngan}, {Ofek},
  {Pogge}, {Santallo}, {Tan}, {Thornley}, {Shin}, {Choi}, {Park}, {Lee}, {Koo},
  \& {{$\mu$}FUN Collaboration}}]{Yee12}
{Yee}, J.~C., {Shvartzvald}, Y., {Gal-Yam}, A., {et~al.} 2012, \apj, 755, 102,
  \dodoi{10.1088/0004-637X/755/2/102}

\bibitem[{{Yee} {et~al.}(2015{\natexlab{a}}){Yee}, {Udalski}, {Calchi Novati},
  {Gould}, {Carey}, {Poleski}, {Gaudi}, {Pogge}, {Skowron}, {Koz{\l}owski},
  {Mr{\'o}z}, {Pietrukowicz}, {Pietrzy{\'n}ski}, {Szyma{\'n}ski},
  {Soszy{\'n}ski}, {Ulaczyk}, \& {Wyrzykowski}}]{Yee15b}
{Yee}, J.~C., {Udalski}, A., {Calchi Novati}, S., {et~al.} 2015{\natexlab{a}},
  \apj, 802, 76, \dodoi{10.1088/0004-637X/802/2/76}

\bibitem[{{Yee} {et~al.}(2015{\natexlab{b}}){Yee}, {Gould}, {Beichman}, {Calchi
  Novati}, {Carey}, {Gaudi}, {Henderson}, {Nataf}, {Penny}, {Shvartzvald}, \&
  {Zhu}}]{Yee15a}
{Yee}, J.~C., {Gould}, A., {Beichman}, C., {et~al.} 2015{\natexlab{b}}, \apj,
  810, 155, \dodoi{10.1088/0004-637X/810/2/155}

\bibitem[{{Yoo} {et~al.}(2004){Yoo}, {DePoy}, {Gal-Yam}, {Gaudi}, {Gould},
  {Han}, {Lipkin}, {Maoz}, {Ofek}, {Park}, {Pogge}, {Mu-Fun Collaboration},
  {Udalski}, {Soszy{\'n}ski}, {Wyrzykowski}, {Kubiak}, {Szyma{\'n}ski},
  {Pietrzy{\'n}ski}, {Szewczyk}, {{\.Z}ebru{\'n}}, \& {OGLE
  Collaboration}}]{Yoo04}
{Yoo}, J., {DePoy}, D.~L., {Gal-Yam}, A., {et~al.} 2004, \apj, 603, 139,
  \dodoi{10.1086/381241}

\bibitem[{{Zhu} {et~al.}(2015){Zhu}, {Udalski}, {Gould}, {Dominik}, {Bozza},
  {Han}, {Yee}, {Calchi Novati}, {Beichman}, {Carey}, {Poleski}, {Skowron},
  {Koz{\l}owski}, {Mr{\'o}z}, {Pietrukowicz}, {Pietrzy{\'n}ski},
  {Szyma{\'n}ski}, {Soszy{\'n}ski}, {Ulaczyk}, {Wyrzykowski}, {OGLE
  Collaboration}, {Gaudi}, {Pogge}, {DePoy}, {Jung}, {Choi}, {Hwang}, {Shin},
  {Park}, {Jeong}, \& {{$\mu$}FUN Collaboration}}]{Zhu15}
{Zhu}, W., {Udalski}, A., {Gould}, A., {et~al.} 2015, \apj, 805, 8,
  \dodoi{10.1088/0004-637X/805/1/8}

\bibitem[{{Zhu} {et~al.}(2016){Zhu}, {Calchi Novati}, {Gould}, {Udalski},
  {Han}, {Shvartzvald}, {Ranc}, {J{\o}rgensen}, {Poleski}, {Bozza}, {Beichman},
  {Bryden}, {Carey}, {Gaudi}, {Henderson}, {Pogge}, {Porritt}, {Wibking},
  {Yee}, {SPITZER Team}, {Pawlak}, {Szyma{\'n}ski}, {Skowron}, {Mr{\'o}z},
  {Koz{\l}owski}, {Wyrzykowski}, {Pietrukowicz}, {Pietrzy{\'n}ski},
  {Soszy{\'n}ski}, {Ulaczyk}, {OGLE Group}, {Choi}, {Park}, {Jung}, {Shin},
  {Albrow}, {Park}, {Kim}, {Lee}, {Cha}, {Kim}, {Lee}, {KMTNET Group},
  {Friedmann}, {Kaspi}, {Maoz}, {WISE Group}, {Hundertmark}, {Street},
  {Tsapras}, {Bramich}, {Cassan}, {Dominik}, {Bachelet}, {Dong}, {Figuera
  Jaimes}, {Horne}, {Mao}, {Menzies}, {Schmidt}, {Snodgrass}, {Steele},
  {Wambsganss}, {RoboNeT Team}, {Skottfelt}, {Andersen}, {Burgdorf}, {Ciceri},
  {D'Ago}, {Evans}, {Gu}, {Hinse}, {Kerins}, {Korhonen}, {Kuffmeier},
  {Mancini}, {Peixinho}, {Popovas}, {Rabus}, {Rahvar}, {Tronsgaard},
  {Scarpetta}, {Southworth}, {Surdej}, {von Essen}, {Wang}, {Wertz}, \&
  {MiNDSTEP Group}}]{Zhu16}
{Zhu}, W., {Calchi Novati}, S., {Gould}, A., {et~al.} 2016, \apj, 825, 60,
  \dodoi{10.3847/0004-637X/825/1/60}

\bibitem[{{Zhu} {et~al.}(2017){Zhu}, {Udalski}, {Calchi Novati}, {Chung},
  {Jung}, {Ryu}, {Shin}, {Gould}, {Lee}, {Albrow}, {Yee}, {Han}, {Hwang},
  {Cha}, {Kim}, {Kim}, {Kim}, {Kim}, {Lee}, {Park}, {Pogge}, {KMTNet
  Collaboration}, {Poleski}, {Mr{\'o}z}, {Pietrukowicz}, {Skowron},
  {Szyma{\'n}ski}, {Koz{\l}owski}, {Ulaczyk}, {Pawlak}, {OGLE Collaboration},
  {Beichman}, {Bryden}, {Carey}, {Fausnaugh}, {Gaudi}, {Henderson},
  {Shvartzvald}, {Wibking}, \& {Spitzer Team}}]{Zhu17a}
{Zhu}, W., {Udalski}, A., {Calchi Novati}, S., {et~al.} 2017, \aj, 154, 210,
  \dodoi{10.3847/1538-3881/aa8ef1}

\end{thebibliography}


\end{document}